\documentclass[amsmath,amssymb,11pt]{article}
\usepackage{jheppub2}
\pdfoutput=1
\usepackage{graphicx,epsfig}
\usepackage{float}
\usepackage{amsmath}
\usepackage{subfloat}
\usepackage[utf8]{inputenc}
\usepackage{hyperref}
\usepackage{caption}
\usepackage{mathrsfs}
\usepackage{subcaption} 
\usepackage{color}
\usepackage{overpic}
\usepackage[dvipsnames]{xcolor}
\usepackage{physics}
\usepackage[multiple]{footmisc}
\usepackage{comment}
\usepackage{tensor}
\usepackage[normalem]{ulem}
\usepackage{orcidlink}

\usepackage{pifont}

\newcommand{\beq}{\begin{equation}}
\newcommand{\eeq}{\end{equation}}

\newcommand{\be}{\begin{equation}}
\newcommand{\ee}{\end{equation}}
\newcommand{\ba}{\begin{eqnarray}}
\newcommand{\ea}{\end{eqnarray}}

\newcommand{\beqa}{\begin{eqnarray}}
\newcommand{\eeqa}{\end{eqnarray}}

\usepackage{tikz}
\usetikzlibrary{positioning}
\usetikzlibrary{intersections}
\usetikzlibrary{fadings} 
\usetikzlibrary{arrows.meta} 
\usetikzlibrary{arrows}

\tikzfading[name=fade out,
inner color=transparent!0,
outer color=transparent!100]

\definecolor{cherryblossompink}{rgb}{1.0, 0.72, 0.77}
\definecolor{lightblue}{rgb}{0.68, 0.85, 0.9}

\usetikzlibrary{decorations.pathmorphing}
\usetikzlibrary{decorations.pathreplacing,decorations.markings}

\usetikzlibrary{backgrounds,automata}

\begin{document}

\title{Quantum Solitons}

\author{Robie A.~Hennigar,}
\author{Ayan K.~Patra,}
\author{and Simon F.~Ross}
\affiliation{Centre for Particle Theory, Department of Mathematical Sciences, Durham University, Durham DH1 3LE, UK}
\emailAdd{robie.a.hennigar@durham.ac.uk, ayan-kumar.patra@durham.ac.uk, s.f.ross@durham.ac.uk}

\abstract{We construct geometries describing the quantum backreaction of thermal fields in AdS$_3$. The solutions are obtained from branes in a four-dimensional AdS C-metric. They can be viewed as solutions of the semiclassical effective theory on the brane, which couples three-dimensional gravity to the CFT dual to the four-dimensional bulk. This brane construction is related by a double analytic continuation to earlier studies of quantum BTZ solutions. There are two families of solutions, labelled by the asymptotic mass. Solutions with negative mass correspond to the back-reaction of a thermal CFT state on global AdS$_3$. Solutions with positive mass have a horizon for zero back-reaction, which is replaced by a smooth origin in the back-reacted solution. We study the thermodynamics and first law on the brane, which we argue is realised in a two-brane setup where we include both the quantum BTZ brane and our quantum soliton brane.}

\maketitle

\section{Introduction}

Semi-classical gravity is a useful approximation to a fully quantum theory, where we study quantum matter fields coupled to classical geometry, considering the effects of the quantum matter on the geometry by treating the expectation value of the stress tensor as a source in Einstein's equations. For a fixed curved spacetime background, it is already challenging to determine the quantum stress tensor even for free fields on this background, and even in cases where it can be determined, the back-reaction of this stress tensor on the geometry is only determined perturbatively. 

Holography offers an alternative approach, which can determine the back-reaction for strongly-coupled conformal field theories that admit holographic duals. We can determine the stress tensor for the fields on a fixed curved background by solving Einstein's equations in the dual theory, reducing the quantum problem to a classical geometrical problem in higher dimensions. Furthermore, if we consider dynamical branes in the dual theory, we can obtain solutions in an effective theory on the brane, which is a coupled theory of classical gravity and the conformal field theory (CFT). The geometry of a brane in the classical bulk can then be interpreted as a solution of the semiclassical equations where we take into account the backreaction of the quantum stress tensor of the conformal field theory. Such constructions thus give insights into the effects of quantum back-reaction in semiclassical gravity.

The main nontrivial example of such a construction is considering branes in the C-metric in four dimensions, giving solutions of semiclassical gravity in the effective three-dimensional theory on the brane \cite{Emparan:1999wa,Emparan:1999fd}---see~\cite{Panella:2024sor} for a recent review. This has been used in particular to construct a quantum version of the BTZ black hole  \cite{Emparan:1999fd,Emparan:2002px,Emparan:2020znc} and various generalizations thereof~\cite{Emparan:2022ijy,Panella:2023lsi, Feng:2024uia, Climent:2024nuj, Climent:2024wol, Bhattacharya:2025tdn}. The geometry on the brane of the quantum BTZ black hole is
\begin{equation} \label{qBTZ}
    ds^2 = -f(r) dt^2 + \frac{dr^2}{f(r)} + r^2 d\phi^2, \quad f(r) = \frac{r^2}{\ell_3^2}-\kappa - \frac{\ell \mu}{r}. 
\end{equation}
This geometry is asymptotically AdS$_3$ with AdS radius $\ell_3$. By choosing the radial coordinate $r$ we can set $\kappa = \pm 1, 0$.\footnote{Note that we have reversed the sign convention for $\kappa$ relative to \cite{Emparan:2020znc}; this is to align with the usual convention for the sign of $\kappa$ in the Schwarzschild-AdS solutions to be discussed in section \ref{SchwAdS}.}    

If we take $\ell/\ell_3 \to 0$ for fixed $\mu$, we recover a locally AdS$_3$ geometry. In this limit, the brane moves to the asymptotic boundary of the spacetime. The bulk solution determines a quantum stress tensor on this boundary geometry. At small but non-zero $\ell$, the effective theory on the brane is three-dimensional Einstein gravity coupled to the holographic CFT, and we can interpret the brane metric \eqref{qBTZ} as a solution of this effective theory, taking into account the back-reaction of the holographic CFT stress tensor. 

The geometry for $\ell=0$ is a BTZ black hole if $\kappa = 1$, with the mass of the black hole encoded in the periodicity of the $\phi$ coordinate. Thus \eqref{qBTZ} can be seen as a quantum-corrected version of the BTZ solution. For $\kappa =-1$, the geometry for $\ell =0$ is a conical defect spacetime.\footnote{$\kappa=0$ is an $M=0$ BTZ black hole; this is a degenerate limiting case, and we will not discuss it in detail in what follows.}
This geometry doesn't have a horizon, but for $\ell \mu >0$, \eqref{qBTZ} has a  horizon for all choices of $\kappa$. Thus, for $\kappa=-1$, the quantum backreaction introduces a more significant qualitative change in the geometry, cloaking the original conical defect singularity in a horizon thereby realizing a form of ``quantum'' cosmic censorship~\cite{Emparan:2002px,Emparan:2020rnp, Kolanowski:2023hvh, Frassino:2025buh}. 

\begin{figure}
    \centering
    \includegraphics[width=0.85\linewidth]{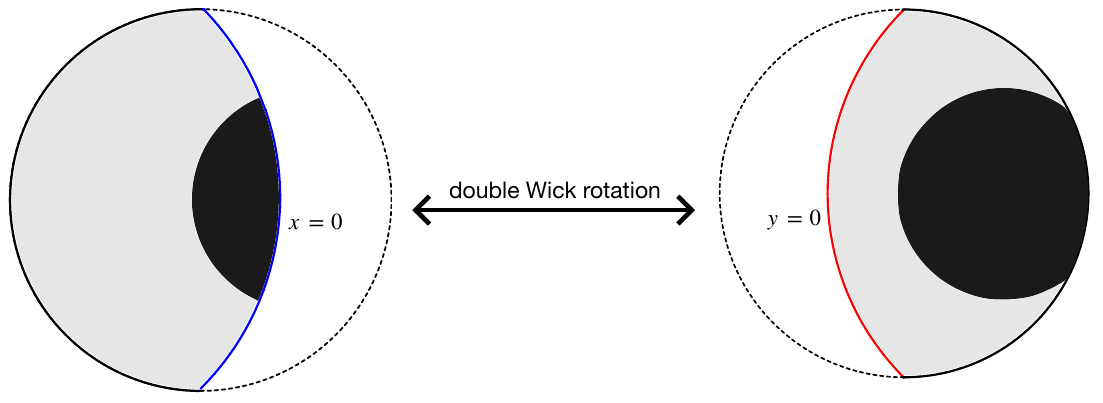}
    \caption{The two choices of brane in the AdS C-metric: on the left, choosing a brane which intersects the black hole horizon gives the quantum BTZ metric \eqref{qBTZ} on the brane. On the right, choosing a brane which does not intersect the horizon gives us the quantum soliton \eqref{qsoliton} on the brane. The two solutions are related by double analytic continuation. The continuation maps the C-metric to itself, but with different parameters, and exchanges the two choices of brane.}
    \label{quSol1}
\end{figure}

The quantum BTZ geometry \eqref{qBTZ} is obtained by considering a particular choice of brane in the C-metric solution. There is another possibility for inserting a brane in this geometry, which was briefly discussed in \cite{Emparan:1999fd} but has not been explored in as much detail. In this paper, we will explore this other choice, which gives a geometry on the brane we dub the \textit{quantum soliton}. The brane metric in this case can be written as
\begin{equation} \label{qsoliton}
     ds^2 = -r^2 d\tau^2 +  \frac{dr^2}{f(r)} + f(r) d\varphi^2,  
\end{equation}
with the same function $f(r)$ as in \eqref{qBTZ}. This solution does not have a horizon; instead there is a circle which shrinks to zero at the zero of $f(r)$. This solution can be obtained from \eqref{qBTZ} by a double analytic continuation, setting $t \to i \varphi$ and $\phi \to i \tau$ (this double analytic continuation was the original inspiration for our work). Our choice of terminology is inspired by the analogy between this and the relation between the Schwarzschild-AdS black hole and the AdS soliton of \cite{Horowitz:1998ha}. In the bulk, this double analytic continuation maps the C-metric to itself, with a different choice of parameters, as we will discuss in section \ref{xycm}. See Figure~\ref{quSol1} for a cartoon illustration of the two different possibilities.

In the limit $\ell/\ell_3 \to 0$, we again recover a locally AdS$_3$ solution. For $\kappa=1$, this is actually global AdS$_3$. The quantum soliton for $\kappa=1$ then describes the back-reaction of a thermal CFT stress tensor, dual to the black hole in the bulk, on global AdS$_3$. For $\kappa=-1$ the quantum soliton corresponds to the back-reaction of a thermal state on a Rindler horizon in AdS$_3$. This case is more interesting: here the solution before back-reaction has a horizon, at $r=0$, but this horizon gets cloaked by a smooth origin for non-zero $\ell$, reversing the pattern seen in the quantum BTZ case.  This is a novel, intrinsically non-perturbative consequence of quantum backreaction.\footnote{This phenomenon was noted earlier in the asymptotically flat case in \cite{Shahbazi-Moghaddam:2025fxi}. We thank Roberto Emparan for bringing this to our attention.} 

In addition to studying the quantum soliton, our paper provides a detailed discussion of the structure of the bulk solutions with these branes. We can consider introducing either the quantum BTZ brane or the quantum soliton brane, or both (as originally considered in \cite{Emparan:1999fd}). We will describe carefully the structure of the bulk in all these cases. 

Before turning to the C-metric solution and its branes, we start in section \ref{SchwAdS} with a discussion of curved geometries on the boundary of an AdS bulk. There is a straightforward story for $\kappa=1$, where we can obtain a BTZ black hole on the boundary of an AdS soliton geometry in a suitable conformal frame \cite{Hubeny:2009rc}, and we can obtain a global AdS$_3$ geometry on the boundary of Schwarzschild-AdS. The stress tensor of the CFT in appropriate states on these backgrounds is then read off from the bulk solutions. For $\kappa=-1$ we explain carefully the global structure of the conformal boundary and relate to previous work in \cite{Czech:2012be,Emparan:2023ypa}. 

We then turn to the constructions with dynamical branes in section \ref{cmetric}. In section \ref{xycm}, we review aspects of the C-metric, and show that the double analytic continuation gives us back the C-metric with different parameters. We review the brane constructions in section \ref{Cbranes}, seeing how we get \eqref{qBTZ} and \eqref{qsoliton} as the induced metric on branes inserted on different surfaces in the C-metric solution. We note that there is an additional horizon in the bulk in some cases which was not apparent in earlier discussions. There are two different limits in which these two branes go off to the asymptotic boundary of the spacetime: in section \ref{Schwlimit} we show that the C-metric can be written exactly as a deformation of Schwarzschild-AdS or the AdS soliton, and discuss the relation to the discussion in section \ref{SchwAdS}. In section \ref{Cboundary} we comment on the structure of the conformal boundary in the C-metric.  

In section \ref{sec:qsol}, we summarize the physical features of the quantum soliton solutions. We first consider $\kappa=1$, showing that these solutions are interpreted as the back-reaction of a quantum stress tensor on global AdS$_3$. For $\kappa=-1$, we discuss how the horizon of the unbackreacted solution obtained in section \ref{SchwAdS} is replaced by a smooth cap.

In section \ref{sec:thermo}, we discuss the thermodynamics of these solutions. A first law for quantum BTZ was obtained in \cite{Emparan:2020znc}. In following a similar logic to obtain a first law for the quantum soliton, we find that a clean bulk analysis leading to a first law requires that we introduce both the quantum soliton brane (at $y=0$ in the coordinates introduced in section \ref{xycm}) and the quantum BTZ brane (at $x=0$). There is a bulk first law in this two-brane setup, which is valid for generic values of the parameters, and gives a first law in the effective theory on the quantum BTZ brane in the limit where its cutoff scale is small, and a first law in the effective theory on the quantum soliton brane in the opposite limit where its cutoff scale is small. We carry out a careful calculation of the Euclidean action in this two-brane setup, taking account of corner terms where the branes meet the cutoff boundary. 

We conclude with a summary of our results and directions for further development in section \ref{sec:disc}.

\section{Unbackreacted solutions}
\label{SchwAdS}

In this section, we review how we can write Schwarzschild-AdS  and its double analytic continuation in coordinates that describe different locally AdS$_3$ geometries on the boundary. We can calculate a stress tensor in the dual CFT on this fixed locally AdS$_3$ geometry from the bulk solution. We can interpret the C-metric solutions that follow in section \ref{cmetric} as describing the back reaction of the CFT stress tensor on the boundary geometry. This construction of locally AdS$_3$ geometries on the boundary was initiated in \cite{Hubeny:2009rc}, who considered BTZ on the boundary of the AdS soliton. Because our interest is in the quantum soliton, we will instead start our discussion from Schwarzschild-AdS. 

The Schwarzschild-AdS$_4$ solution is 
\be \label{SAdS}
\dd s^2 = -F(\rho)\ell_4^2 \dd \tau^2 + \frac{\dd \rho^2}{F(\rho)} + \rho^2 d\Omega_{\kappa}^2\, ,
\ee
with
\be \label{SchwF}
F(\rho) = \frac{\rho^2}{\ell_4^2} + \kappa - \frac{2 G_4 M}{\rho} \, ,
\ee
where we have taken a factor of $\ell_4$ out to make the coordinate $\tau$ dimensionless for later convenience, and $d\Omega_{\kappa}^2$ is the round metric on $S^2$ for $\kappa=1$, a flat metric for $\kappa=0$, and a locally $H^2$ space for $\kappa=-1$.\footnote{The $\kappa$ here will be identified with the $\kappa$ introduced in the introduction; this is responsible for the choice of sign of $\kappa$ there.} There is a horizon in the bulk at $\rho = \rho_0$, where $F(\rho_0)=0$. The black hole in the bulk corresponds to a thermal state in the CFT, at temperature 
\begin{equation} \label{temp}
    T = \frac{F'(\rho_0)}{4\pi} = \frac{3 \rho_0^2 + \kappa \ell_4^2}{4 \pi \ell_4^2 \rho_0}.
\end{equation}

\subsection{Thermal AdS$_3$ and the BTZ black hole}

Consider first the case $\kappa=1$. Then in standard coordinates 
\begin{equation} \label{sphere}
    d\Omega_{\kappa}^2 = d\theta^2 + \sin^2 \theta d\varphi^2. 
\end{equation}
The boundary is the Einstein static universe (ESU) $S^2 \times \mathbb{R}$, with geometry 
\begin{equation} \label{ESU}
 ds_3^2 = - d \tau^2 + d\theta^2 + \sin^2 \theta d\varphi^2.
\end{equation}
The boundary stress tensor reads
\begin{equation} \label{CFTstress}
    \langle T_{\mu}^{\ \nu} \rangle = \frac{M \ell_4}{8\pi} {\rm diag}\left(-2,1,1  \right) .
\end{equation}
The form of the stress tensor is fixed by symmetry, together with tracelessness: the spatial part of the stress tensor is fixed to be a constant multiple of the identity by spherical symmetry, and the time component is then fixed by tracelessness. 

We can rewrite the  half of the boundary with $\theta \in (0, \pi /2)$ as global AdS$_3$,\footnote{Note that we naturally get global AdS$_3$, and not a conical defect, as the metric on $S^2$ has $\Delta \varphi = 2\pi$.} by writing $\sin \theta = \tanh \gamma$ with $\gamma \in (0, \infty)$ and rescaling the boundary by a conformal factor of $\cosh^2\gamma$, to obtain a boundary metric
\begin{equation} \label{gAdSb}
 ds_3^2 = - \cosh^2 \gamma d \tau^2 + d\gamma^2 + \sinh^2 \gamma d\varphi^2.
\end{equation}
The other half of the boundary can similarly be mapped to a second copy of AdS$_3$. Thus, the bulk Schwarzschild-AdS geometry can also be interpreted as dual to a CFT on two copies of global AdS$_3$, with the conformal boundaries identified, with stress tensor
\begin{equation} \label{Bstress}
    \langle T_{\mu}^{\ \nu} \rangle = \frac{M \ell_4}{8\pi \cosh^3 \gamma} {\rm diag}\left(-2,1,1  \right) .
\end{equation}
The quantum solitons that we obtain later describe the back-reaction of this thermal stress tensor on global AdS$_3$.

If we perform the double analytic continuation $\tau \to i \phi$, $\varphi \to it$ on Schwarzschild-AdS, we obtain an AdS soliton, 
\be \label{adssol}
\dd s^2 = F(\rho)\ell_4^2 \dd \phi^2 + \frac{\dd \rho^2}{F(\rho)} + \rho^2 (d\theta^2 - \sin^2 \theta dt^2)\, .
\ee
This has a smooth origin at $\rho=\rho_+$ where $F(\rho_+)=0$.  This is a bubble of nothing with a dS$_2 \times S^1$ boundary.\footnote{Higher-dimensional versions of this bulk solution were discussed  in \cite{Balasubramanian:2002am}.} Doing this analytic continuation in the conformal frame where the Schwarzschild-AdS boundary is global AdS$_3$ gives us a BTZ black hole on a portion of the boundary of the soliton, 
\begin{equation} \label{BTZd}
 ds_3^2 = - \sinh^2 \gamma dt^2 + d\gamma^2 + \cosh^2 \gamma d\phi^2,
\end{equation}
recovering the construction of \cite{Hubeny:2009rc}. The BTZ horizon corresponds to the horizon of the static patch in the dS$_2 \times S^1$ conformal frame. The boundary of BTZ is at $\theta=\pi/2$, which is the worldline of an observer at the center of the static patch.  If we write $\Delta \phi = 2 \pi r_+$, we can cast this in a standard BTZ form by writing $\phi = r_+ \bar \phi$, $t = r_+ \bar t$, and $\cosh \gamma = \frac{r}{r_+}$. The boundary metric is then 
\begin{equation}
 ds_3^2 = - (r^2 - r_+^2) d\bar t^2 + \frac{dr^2}{r^2 - r_+^2} + r^2 d\bar \phi^2. 
\end{equation}
The CFT stress tensor in BTZ coordinates is the analytic continuation of \eqref{Bstress},
\begin{equation} \label{BTZstress}
    \langle T_{\mu}^{\ \nu} \rangle = \frac{M \ell_4}{8\pi \cosh^3 \gamma} {\rm diag}\left(1,1,-2  \right) .
\end{equation}
The quantum BTZ solutions 

The interpretation of the CFT stress tensor is however somewhat different in these two cases. For the Schwarzschild-AdS case, $M$ is a parameter labeling the state. Thus, we have a one-parameter family of thermal CFT states on global AdS$_3$, with stress tensor \eqref{Bstress}. In the AdS soliton, by contrast, smoothness of the bulk geometry fixes the periodicity of $\phi$ to be 
\begin{equation} \label{phiFper}
    \Delta \phi = \frac{4\pi}{\ell_4 F'(\rho_0)} = \frac{4\pi \ell_4 \rho_0}{3\rho_0^2 + \ell_4^2}. 
\end{equation}
For a given choice of period $\Delta \phi$, there are two solutions for $\rho_0$ and hence for $M$, while the $M=0$ solution is locally AdS$_4$ and is compatible with any period for $\phi$. Thus, for a given boundary geometry, there are three possible bulk solutions, corresponding to three possible states of the CFT on this background. For $M=0$ the boundary stress tensor vanishes. The two states with non-zero $M$ give rise to the two branches of quantum BTZ solutions found in \cite{Emparan:1999fd} when we turn on back-reaction. 

In \cite{Emparan:2020znc}, \eqref{BTZstress} was obtained as the leading part of the CFT stress tensor in the quantum BTZ geometry. It was compared to calculations of the stress tensor in free theories on the BTZ background with transparent boundary conditions, and it was noted that the form of the stress tensor matched. This matching can be understood more simply from the above perspective. While in the BTZ coordinates \eqref{BTZd} the structure of the stress tensor is not obviously dictated by symmetry, for a CFT with transparent boundary conditions it is related by a conformal transformation to a stress tensor on dS$_2 \times S^1$ (and by double analytic continuation to the thermal stress tensor on the ESU). For de-Sitter invariant states the form of the stress tensor in this conformal frame {\it is} fixed by symmetry (by analogy to the argument we reviewed above for the ESU). So it is not surprising that the results for free theories and the holographic results match. 

The assumption of transparent boundary conditions is however crucial in this argument. In the global AdS$_3$ case, the stress tensor for fields with Dirichlet boundary conditions was recently calculated in \cite{Thompson:2025jkn,Thompson:2025kfm}, and the form of the stress tensor obtained there doesn't match the structure above. 

\subsection{Rindler horizons and conical defects in AdS$_3$}

Consider now $\kappa=-1$. The Schwarzschild-AdS black hole now has a rather more complicated structure.  In this case the mass parameter $M$ can be negative, $G_4 M \geq - 3^{-3/2} \ell_4$, and the two dimensional spatial metric in \eqref{SAdS} is a hyperbolic disc. To consider these as black hole solutions we usually take a quotient of $H^2$ so that the horizon becomes a compact Riemann surface $\Sigma_g$ with a negatively curved metric. The full eternal black hole spacetime then has two asymptotic regions with boundaries $\Sigma_g \times \mathbb R$. This is the hyperbolic black hole discussed in \cite{Emparan:1999gf}.

For the present purpose however we want to consider a situation where the two-dimensional metric is the full hyperbolic disc. The bulk spacetime still looks like it has two asymptotic regions, but the boundary in each is now $H^2 \times \mathbb R$, so the boundary has a spatial boundary, and we should understand what happens there. This was previously considered in \cite{Czech:2012be,Emparan:2023ypa}, who argued that the boundaries of the two hyperbolic discs are identified, so the bulk solution is dual to a CFT on a single space which is split by a Rindler horizon.  

This is obvious for the case with $M=0$. The bulk geometry \eqref{SAdS} is then AdS$_4$ in a Rindler-like coordinate system, and the bulk horizon is just an acceleration horizon. The asymptotic boundary of AdS$_4$ is $S^2 \times \mathbb R$, which is conformally flat,  so we can conformally map it to Minkowski space such that the bulk horizon in the coordinates of \eqref{SAdS} meets the conformal boundary along the Rindler horizon of the Minkowski coordinates. This Rindler horizon splits the $S^2$ into two hyperbolic discs. The coordinates of \eqref{SAdS} cover one side of the Rindler horizon on the boundary, and the two $H^2$s are identified along their asymptotic boundaries to obtain the $S^2$ in global coordinates. 

We would expect moving to non-zero mass to modify the state, rather than the geometry the field theory lives on, so the bulk black hole for non-zero mass should still be dual to a state in the field theory on two copies of $H^2$ with their asymptotic boundaries identified. This can be seen most cleanly by considering the behaviour of two-point functions with one operator insertion on each conformal boundary. In the coordinates of \eqref{SAdS}, if we consider the correlator between an operator on the left boundary and an operator on the right boundary at the same value of $t$ and the same position on $H^2$, by symmetry the result is independent of $t$ and the point on $H^2$, but it may be a function of $M$, $\langle O_L O_R \rangle = C(M)$. The mapping from $H^2$ to a hemisphere of $S^2$ involves a conformal factor $\Omega$ which vanishes at the equator of the $S^2$. Thus, in the $S^2$ conformal frame, the correlator is $\langle O_L O_R \rangle = C(M) \Omega^{-2\Delta}$, and it has a divergence as the points on the two hemispheres approach the equator.  This divergence can be interpreted as a short distance singularity, signaling that the equators of the two hemispheres are identified. 

The change in the state as we change $M$ is signalled by the $M$ dependence of the two-point function. In the $M=0$ case, the coefficient of the divergence is canonical, and the state is smooth on the equator of $S^2$; indeed this is just the vacuum state of the CFT in global coordinates. For $M \neq 0$, the coefficient of the divergence in the two-point function on the equator is different, so the state is not smooth there. This non-smoothness of the state can also be seen by considering the CFT stress tensor on the boundary. The CFT dual to \eqref{SAdS} is in a thermal state at the temperature \eqref{temp}. As in the $\kappa=1$ case, symmetry fixes the boundary stress tensor in the $H^2$ conformal frame to be 
\begin{equation} 
    \langle T_{\mu}^{\ \nu} \rangle = \frac{M \ell_4}{8\pi} {\rm diag}\left(-2,1,1  \right) .
\end{equation}
If we rescale from $H^2$ to $S^2$, the stress tensor on $S^2$ is 
\begin{equation} 
    \langle T_{\mu}^{\ \nu} \rangle = \frac{M \ell_4}{8\pi \Omega^3} {\rm diag}\left(-2,1,1  \right) ,
\end{equation}
which clearly diverges for non-zero mass on the equator of the $S^2$ where $\Omega \to 0$. Note that $M=0$ has vanishing stress tensor, but not vanishing temperature; the temperature is given by \eqref{temp}, which for $M=0$ and $\kappa =-1$ gives $T = \frac{1}{2\pi \ell_4}$, which is precisely the Rindler temperature of the bulk and boundary horizons. For non-zero $M$, the temperature changes; positive $M$ increases the temperature while negative $M$ decreases it, reaching zero temperature at $G_4 M = - 3^{-3/2} \ell_4$.\footnote{These geometries were previously discussed in \cite{Czech:2012be,Emparan:2023ypa} from the perspective of considering how changing this temperature corresponds to changing the entanglement between the CFT on the two $H^2$s, and the relation to the length of the wormhole in the bulk. In \cite{Emparan:2023ypa}, the focus was on $M<0$, exploring how the Einstein-Rosen bridge gets longer as we reduce the entanglement. We will see below that we can only address the case with $M>0$. }  Hence, in the spirit of the Kay-Wald theorem~\cite{Kay:1988mu}, the singularity of the stress tensor is a consequence of putting a thermal state at the wrong temperature on a spacetime with a horizon.\footnote{The quantum state of bulk fields is regular on the bulk horizon. But in the boundary, we have the CFT on a fixed geometry at different temperatures.}  The quantum soliton will describe the back-reaction of this stress tensor on this background. Since the state is singular on the horizon, this back-reaction is never negligible.

We now consider the coordinates on $H^2$. Our aim is to rewrite it in a locally AdS$_3$ form, to make contact with the quantum BTZ and quantum soliton solutions to follow.  It is therefore convenient to introduce a non-standard set of coordinates on $H^2$, where we write 
\begin{equation} \label{H2}
    d\Omega_{\kappa}^2 = d\theta^2 + \cosh^2 \theta d\varphi^2, 
\end{equation}
with $\theta \in (0, \infty)$. These are related to the embedding coordinates $T,X,Y$ where $H^2$ is $-T^2+X^2+Y^2=-1$ by 
\begin{equation}
    T = \cosh \theta \cosh \varphi, \quad X = \cosh \theta \sinh \varphi, \quad Y = \sinh \theta, 
\end{equation}
so $\varphi$ is a hyperbolic angle, with no periodic identification, and $\theta \geq 0$ covers half the hyperbolic disc, with $Y \geq 0$. The boundary metric is 
\begin{equation} \label{kminusb}
 ds_3^2 = - d\tau^2 + d\theta^2 + \cosh^2 \theta d\varphi^2.
\end{equation}
We can conformally rescale this to a locally AdS$_3$ metric by setting  $\cosh \theta = \coth \gamma$ for $\gamma \in (0, \infty)$ (where $\gamma =0$ corresponds to $\theta = \infty$ and vice-versa), and rescaling by a conformal factor of $\sinh^2 \gamma$. Then the boundary metric is 
\begin{equation} \label{BTZb2}
 ds_3^2 = - \sinh^2 \gamma d\tau^2 + d\gamma^2 + \cosh^2 \gamma d\varphi^2.
\end{equation}
This is AdS$_3$ written in BTZ coordinates; however the angle $\varphi$ is not periodically identified. This metric thus covers half of each of the two $H^2$s on the boundary of the bulk solution, with the horizon at $\gamma=0$ forming part of the Rindler-like horizon separating the two $H^2$s. In these coordinates, the boundary stress tensor is 
\begin{equation} 
    \langle T_{\mu}^{\ \nu} \rangle = \frac{M\ell_4}{8\pi \sinh^3 \gamma} {\rm diag}\left(-2,1,1  \right) .
\end{equation}
In the next section, we will see how the geometry on the brane in the C-metric solutions gives a back-reacted version of this solution. There are subtleties in this case in the limit where we remove the back-reaction, as we will see. 

This is the case that will give us the quantum soliton. To get the unbackreacted version of quantum BTZ, we can consider the same analytic continuation as before, sending $\tau \to i \phi$, $\varphi \to it$. This maps the $\kappa=-1$ bulk to 
\be \label{hypsol}
\dd s^2 = F(\rho)\ell_4^2 \dd \phi^2 + \frac{\dd \rho^2}{F(\rho)} + \rho^2 (d\theta^2 - \cosh^2 \theta dt^2)\, .
\ee
This is again an AdS soliton geometry; the geometry closes off smoothly in the interior at $\rho_0$ where $F(\rho_0)=0$, fixing the periodicity $\Delta\phi$ of the angular coordinate. In this case the boundary is AdS$_2 \times S^1$.\footnote{The unusual choice of coordinates on $H^2$ is motivated by the fact that the analytic continuation gives us global coordinates on the AdS$_2$ (although the coordinate range $\theta \in (0, \infty)$ taken above only covers half the AdS$_2$ space).}  The case with $M=0$ is again AdS$_4$, but unlike in the black hole case this coordinate system covers the whole of AdS$_4$. In particular $t$ is identified with the time coordinate in global coordinates on AdS$_4$. Thus in this case there are no subtleties analogous to our discussion in the black hole case. Setting $\cosh \theta = \coth \gamma$ and rescaling by a factor of $\sinh^2 \gamma$ maps the conformal boundary to a locally AdS$_3$ form,
\begin{equation}
 ds_3^2 = - \cosh^2 \gamma dt^2 + d\gamma^2 + \sinh^2 \gamma d\phi^2.
\end{equation}
 In the AdS$_2 \times S^1$ form of the boundary, $\gamma=0$ corresponds to one of the asymptotic boundaries of AdS$_2$.  These are global AdS$_3$ coordinates, but because the bulk solution fixes
\begin{equation} 
    \Delta \phi = \frac{4\pi}{\ell_4 F'(\rho_0)} = \frac{4\pi \ell_4 \rho_0}{3\rho_0^2 - \ell_4^2},
\end{equation}
the boundary metric is not smooth at $\gamma=0$. It is a conical defect for $M>0$, and a conical excess for $M <0$.\footnote{This is related to the observation above that the state of the fields on the conformal boundary is not smooth on the Rindler horizon separating the two $H^2$ patches; the conical defect appearing here is the Euclidean manifestation of this mismatch of temperatures.}
Unlike for $\kappa =1$, here there's a unique positive solution for $\rho_0$ for given $\Delta \phi$, so we only get one CFT state for each given boundary geometry. The boundary stress tensor is 
\begin{equation}
    \langle T_{\mu}^{\ \nu} \rangle = \frac{M\ell_4}{8\pi\sinh^3 \gamma} {\rm diag}\left(1,1,-2  \right) .
\end{equation}
This is not smooth at the origin $\gamma=0$ in the BTZ conformal frame. The quantum BTZ solutions describe the back-reaction of this stress tensor on the geometry for $M>0$, which hides the singular origin behind a horizon.

\section{C-metric and brane configurations}
\label{cmetric}

We now turn to the construction of back-reacted versions of these geometries, where we couple the CFT stress tensor to dynamical gravity. These back-reacted geometries are obtained as the worldvolume geometry on constant tension branes in the AdS C-metric, so we now discuss in detail the structure of the AdS C-metric and the different constant tension branes we can consider in it.  We write the C-metric as 
\be \label{xycm}
\dd s^2 = \frac{1}{A^2 (x-y)^2} \left[H(y) \dd t^2 - \frac{\dd y^2}{H(y)} + \frac{\dd x^2}{G(x)} + G(x) \dd \phi^2 \right] \, ,
\ee
where 
\be 
G(x) = 1 - k x^2 - \mu x^3 \, , \quad H(y) = - \left(\lambda + k y^2 + \mu y^3 \right) \, .
\ee
This is essentially the description used in \cite{Emparan:1999fd}, but with some minor differences in notation. We take the coordinates to be dimensionless, so the constants appearing in $G,H$ are also dimensionless. The overall constant $A$ has dimensions of inverse length. 
This is an asymptotically AdS$_4$ solution of the Einstein equations with a negative cosmological constant, where the AdS length scale is 
\be  \label{l4}
\ell_4 = \frac{1}{A \sqrt{\lambda + 1}} \, .
\ee
By scaling the coordinates we can set $k = 0, \pm 1$. The asymptotic boundary of the spacetime is at $y=x$, and without loss of generality we can take $y \leq x$.

The quantum soliton \eqref{qsoliton} and the quantum BTZ solution \eqref{qBTZ} are related by a double analytic continuation. In the previous section, the bulk solutions were similarly related by a double analytic continuation. In the C-metric, for $\lambda >0$, the same analytic continuation, sending $t \to i \varphi$, $\phi \to i \tau$ just gives us back another C-metric with a different set of parameters. After analytically continuing, defining 
\begin{equation} \label{contxy}
    x = -\sqrt{\lambda} \hat{y} \, , \quad y = -\sqrt{\lambda} \hat{x} \, , \quad \tau = \frac{\hat{t}}{\sqrt{\lambda}}\,, \quad \varphi = \frac{\hat{\phi}}{\sqrt{\lambda}} \, ,
\end{equation}
and 
\begin{equation} \label{contpar}
     \hat{A}=A \sqrt{\lambda} ,\quad \hat{\lambda} = \frac{1}{\lambda} \, ,  \quad \hat{k} = - k \, , \quad \hat{\mu} =  \mu \sqrt{\lambda}\, ,
\end{equation}
we recover the same metric \eqref{xycm} in the hatted coordinates and parameters. Thus, we will be able to obtain the quantum soliton and quantum BTZ as geometries on different branes in the C-metric.

\subsection{C-metric root structures}

The C-metric will have different structures depending on the roots of $G(x)$ and $H(y)$. These are cubics in $x,y$, so they have at most three roots. Since $H(\xi) = G(\xi) - (\lambda +1)$, the graph of $H$ always lies below the graph of $G$ for the same value of the argument. $H'(\xi) = G'(\xi) = -2k \xi - 3 \mu \xi^2$, so both functions always have two turning points, at $\xi=0$ and $\xi = \xi_c = -\frac{2k}{3\mu}$.

We will restrict to $\mu >0$ and $\lambda >0$. With $\mu <0$, the bulk spacetime has naked singularities, which are not removed by considering branes, so these solutions are not physically interesting.\footnote{The C-metric has a symmetry under $(x, y, \mu) \to - (x, y, \mu)$, so cases with $\mu <0$ can be obtained from the cases with $\mu >0$ above by this inversion. Considering the region with $y \leq x$ in the $\mu <0$ cases then corresponds to considering the region with $y \geq x$ in the $\mu >0$ cases obtained under this inversion.} 
In the C-metric, we can consider $-1 < \lambda <0$ (we need $1+\lambda >0$ to hold $\ell_4$ in \eqref{l4} fixed). However, we will see below that to have an AdS metric on the brane of the form we are interested in when we insert a brane we need $\lambda >0$, so we will not describe the cases with $\lambda <0$ in detail.

\begin{figure}
    \centering
    \includegraphics[width=\linewidth]{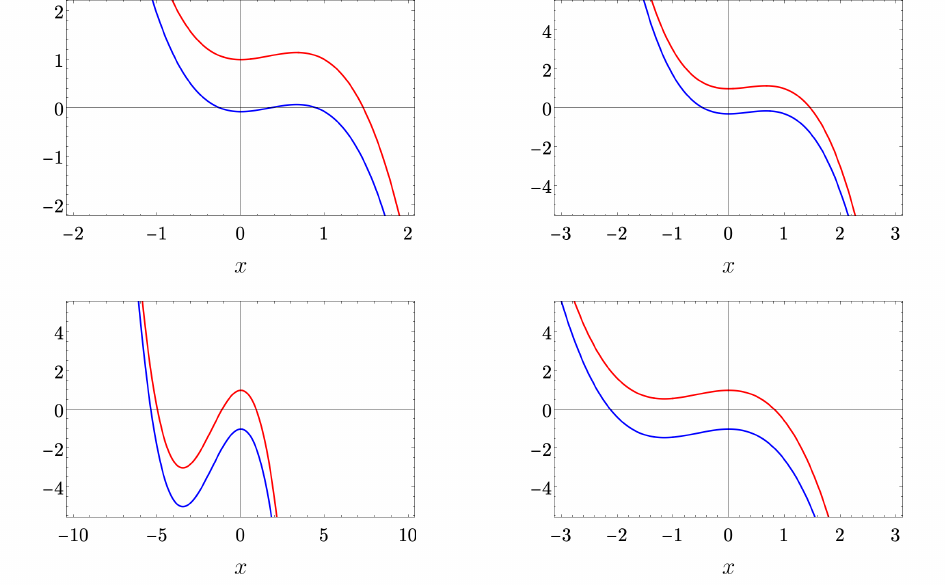}
    \caption{Plots of the different root configurations that the functions $H$ and $G$ can have. Here, the blue curves are $H(\xi)$ while the red curves are $G(\xi)$. The plots are: \textbf{Case I} (top left), \textbf{Case II} (top right), \textbf{Case III} (bottom left), \textbf{Case IV} (bottom right). }
    \label{fig:G_H_plots}
\end{figure}

For $\mu >0$ and $\lambda >0$, $G(x)$ will always have one positive root, which we call $x=x_+$, and $H(y)$ will have one negative root, which we call $y=y_-$. There are four qualitatively different cases for other roots:\footnote{There are also cases where $G$ or $H$ has a double root. We shall not consider those in detail here.}  
\begin{itemize}
    \item {\bf Case I}: $k = -1$,  $0 < \lambda < 4/(27 \mu^2)$. Here $G(x)$ has just one real root, but $H(y)$ has three real roots, the universal negative one at $y = y_-$ and two additional positive roots. 
    \item {\bf Case II}: $k = -1$, $\lambda > 4/(27 \mu^2)$. Here both $G(x)$ and $H(y)$ have only one real root. 
    \item {\bf Case III}: $k = +1$, $ \mu < 2/(3 \sqrt{3})$. Here $G(x)$ has three real roots, the universal positive one at $x=x_+$ and two more negative ones, while $H(y)$ has just one real root. 
    \item {\bf Case IV}: $k = +1$, $\mu > 2/(3 \sqrt{3})$. Here both $G(x)$ and $H(y)$ have only one real root. 
\end{itemize}
The double analytic continuation \eqref{contpar} exchanges case I and case III, and case II and case IV. The shapes of $G(x)$ and $H(x)$ in these cases can be seen in Figure~\ref{fig:G_H_plots}. We usually restrict to a region with $G>0$ and $H <0$, where $\phi$ is spacelike and $t$ is timelike. Zeros of $G(x)$ are axes, where the $\phi$ circle degenerates, and zeros of $H$ are horizons, where the $t$ direction becomes null.  We will always choose the periodicity of $\phi$ to make the metric smooth at $x=x_+$, which requires
\begin{equation} \label{phiCper}
    \Delta \phi = \frac{4\pi}{|G'(x_+)|} = \frac{4\pi x_+}{3 - k x_+^2}.
\end{equation}

\begin{figure}
    \centering
    \includegraphics[width=0.45\linewidth]{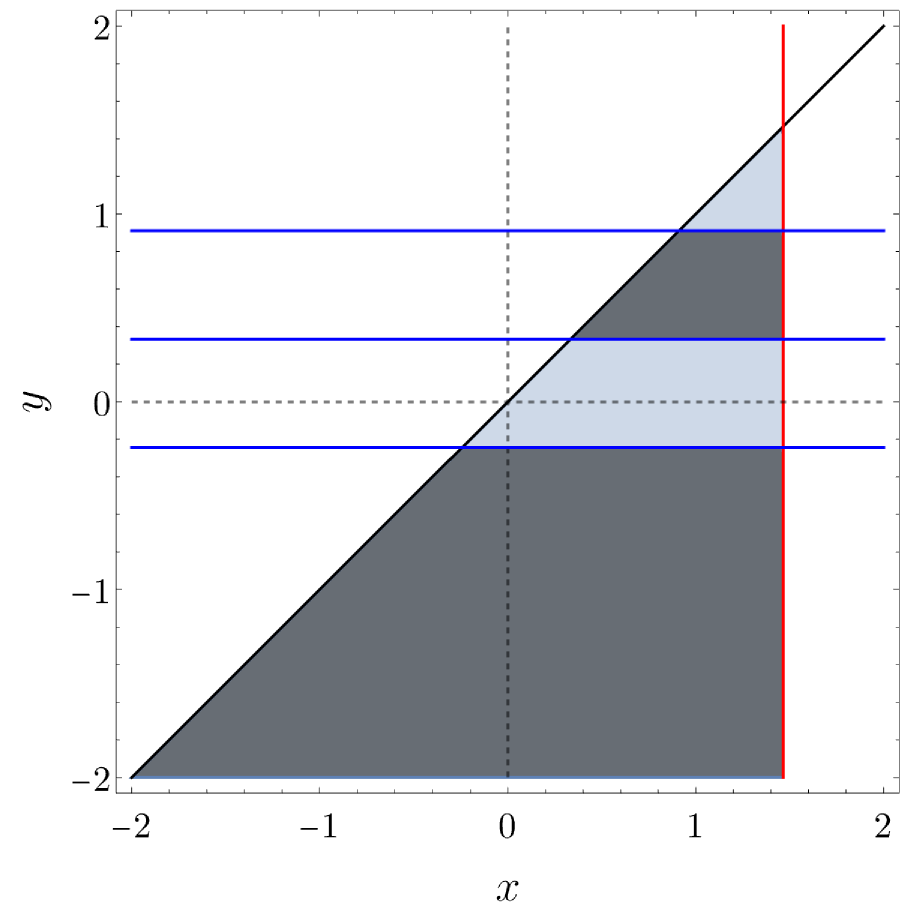}
    \includegraphics[width=0.45\linewidth]{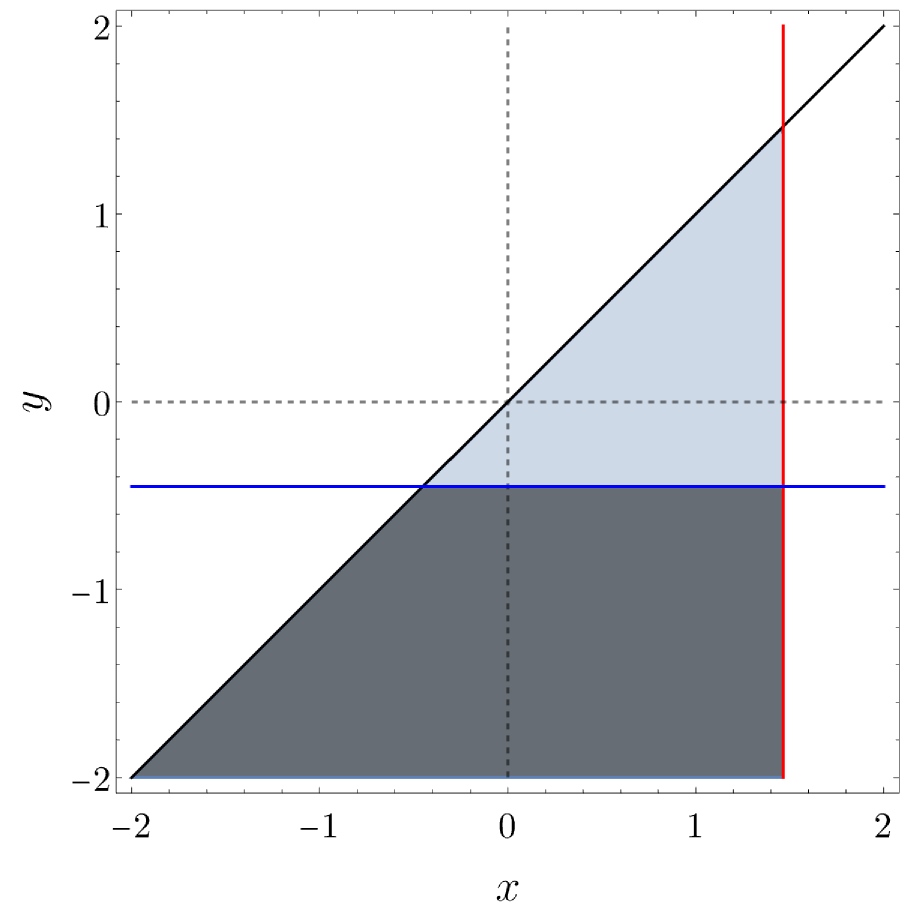}
    \includegraphics[width=0.45\linewidth]{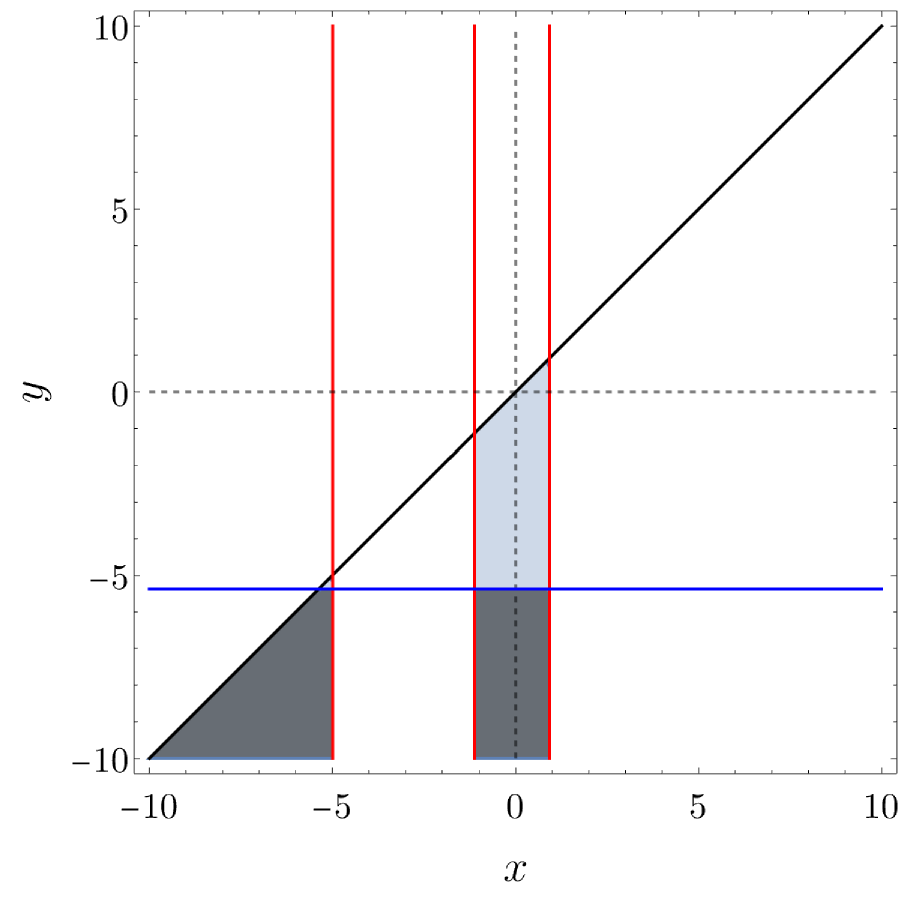}
    \includegraphics[width=0.45\linewidth]{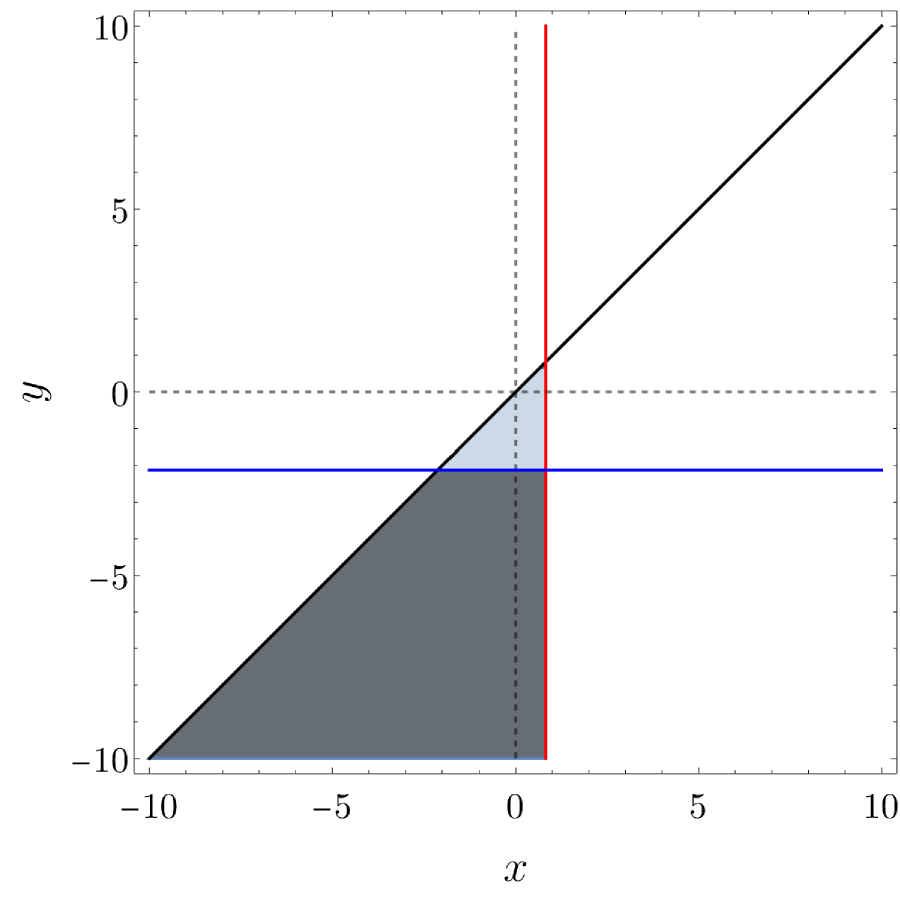}
    \caption{Parameter space in $(x,y)$ plane for:  \textbf{Case I} with $\kappa = -1$, $\mu = 1$ and $\lambda = 2/27$ (top left), \textbf{Case II} with  $\kappa = -1, \mu = 1, \lambda = 8/27$ (top right), \textbf{Case III} with  $\kappa = +1$, $\mu = 1/(3\sqrt{3})$, $\lambda = 1$ (bottom left), and \textbf{Case IV} with $\mu = 1/(\sqrt{3})$, $\lambda = 1$ (bottom right). In all cases, the solid black line $y = x$ is the asymptotic boundary. The red lines are the zeros of $G(x)$, and the blue lines are the zeros of $H(y)$. The dashed gray lines are possible locations of branes. The shaded regions are parts of the relevant spacetime: The light blue shading corresponds to $H(y) < 0$, corresponding to untrapped regions, while the black shaded regions are where $H(y) > 0$, which correspond to trapped regions. If a region is white it is not relevant, either because it's on the ``other side'' of the conformal boundary, or because $G(x) < 0$ there.}
    \label{fig:xy_diagrams}
\end{figure}

In considering the C-metric on its own, we usually restrict attention to case III, where $G(x)$ has three roots, and take $x$ to lie between the two larger roots, $x \in (x_-, x_+)$, where $G(x)$ is positive. The horizon at $y = y_-$ is then a compact black hole horizon, and there is a conical defect along the axis at  $x=x_-$.\footnote{Unlike in the flat space C-metric, in the AdS C-metric when the black hole horizon is a sphere there is no additional acceleration horizon in the spacetime.} When we consider inserting branes, it can also be interesting to consider other cases, as we discuss below. It would be interesting to understand the global structure of the spacetime in case I.\footnote{One interesting feature is that there is a region of trapped surfaces between the two additional horizons which includes a portion of the asymptotic boundary.}

\subsection{Branes in the C-metric}
\label{Cbranes}

We want to consider inserting a braneworld into this C-metric solution. We consider constant-tension branes, so the branes can be inserted along umbilic submanifolds, where the extrinsic curvature is proportional to the induced metric on the surface, $K_{ab} \propto h_{ab}$. If we define the length scale $\ell$ associated with the brane tension by setting 
\begin{equation}
    K_{ab} = \frac{1}{\ell} h_{ab}, 
\end{equation}
in the induced theory on the brane we will have a curvature scale $\ell_3$, where
\begin{equation} \label{l4l3l}
    \frac{1}{\ell_4^2} = \frac{1}{\ell^2} + \frac{1}{\ell_3^2}. 
\end{equation}

It was shown by Karch and Randall that a massive graviton state localizes on the brane~\cite{Karch:2000ct}. Hence, there is an induced theory of dynamical three-dimensional gravity residing on the brane. Holographically, the theory living on the brane can be obtained by `integrating out' CFT degrees of freedom above a cutoff energy scale $\sim 1/\ell$~\cite{deHaro:2000wj,Emparan:1999pm}. The resulting effective action for the theory on the brane is~\cite{Chen:2020uac,Emparan:2020znc}
\be \label{eqn:brane_action}
I_{\rm brane} = \frac{1}{16 \pi G_3} \int \dd^3 x \sqrt{-h} \left[\frac{2}{L_3^2} + R + \ell^2 \left(\frac{3}{8}R^2 - R_{ab}R^{ab} \right) +\cdots  \right] + I_{\rm CFT} \, ,
\ee
where 
\be 
G_3 = \frac{G_4}{2 \ell_4}  \qquad \text{and} \qquad \frac{1}{L_3^2} = \frac{1}{\ell_3^2}\left(1 + \frac{\ell^2}{4 \ell_3^2} \right) \, .
\ee
The factor of $2$ in the definition of the three-dimensional bare Newton constant arises because we consider the branes to be two-sided. The difference of $L_3$ from the curvature of the brane is due to the latter receiving a correction from higher-curvature terms in the brane effective action. Thus, from the brane perspective $\ell_3$ is the three-dimensional AdS scale and $\ell$ is a UV scale at which higher-derivative corrections become important. Similarly, the three-dimensional Newton's constant receives corrections from the higher-curvature terms, with the ``renormalized'' Newton's constant reading~\cite{Emparan:2020znc}
\be \label{eqn:calG3}
\mathcal{G}_3 = \frac{\ell_4}{\ell} G_3 = \frac{G_4}{2 \ell} \, .
\ee
The effective action above is an expansion in powers of $\ell$, so it is a useful description in the parameter region $\ell/{\ell_3} \ll 1$.

%
%
%
%
%
%

In the C-metric, the umbilic submanifolds where we can insert a brane are at constant $x$ or $y$, at the turning points of $G$, at $x=0$, $x=\xi_c=-\frac{2k}{3\mu}$, $y=0$, or $y = \xi_c$. If we consider inserting a single brane, we can restrict attention to inserting it at $x=0$ or $y=0$, as considering a brane at $x=\xi_c$ or $y=\xi_c$ is equivalent to inserting a brane at $x=0$ or $y=0$ in a C-metric with different parameters \cite{Emparan:1999fd}. We can also consider inserting both a brane at constant $x$ and a brane at constant $y$: for these two-brane scenarios there are genuinely different possibilities, where we insert a brane at $x=0$ and $y=0$, at $x=0$ and $y=\xi_c$, or $x=\xi_c$ and $y=0$. 

\subsubsection{$x=0$ brane: quantum BTZ}

 Consider first inserting a brane at $x=0$, which is the case usually discussed. The extrinsic curvature is $K_{ab} = -A h_{ab}$. Inserting a positive-tension brane will restrict us to the region $x >0$, and we have 
\begin{equation} \label{ll3x}
\ell = \frac{1}{A} \quad \Rightarrow \quad  \ell_3 = \frac{1}{A \sqrt{\lambda}}.     
\end{equation}
%

%
We see that we need $\lambda >0$ for real $\ell_3$; taking $\lambda <0$ would give a de Sitter geometry on the brane~\cite{Emparan:2022ijy}.  We get the quantum BTZ geometry \eqref{qBTZ} on the worldvolume of this brane \cite{Emparan:2002px,Emparan:2020znc}. Explicitly, the induced metric on the brane is 
\begin{equation}
    ds_3^2 = \frac{1}{A^2 y^2} \left[ H(y) dt^2 - \frac{dy^2}{H(y)} + d\phi^2 \right], 
\end{equation}
and setting 
\begin{equation} \label{branec}
    r = -\frac{1}{Ay}, \quad t = A \tilde t,
\end{equation}
we get the metric \eqref{qBTZ} on the brane in $\tilde t, r, \phi$ coordinates, with $f(r) = -\frac{H(y)}{y^2}$ and $\kappa=-k$. Note that on the $x=0$ brane, $y \leq 0$, so $r \in (0, \infty)$ covers the whole brane. 

\begin{figure}
    \centering
    \includegraphics[width=0.85\linewidth]{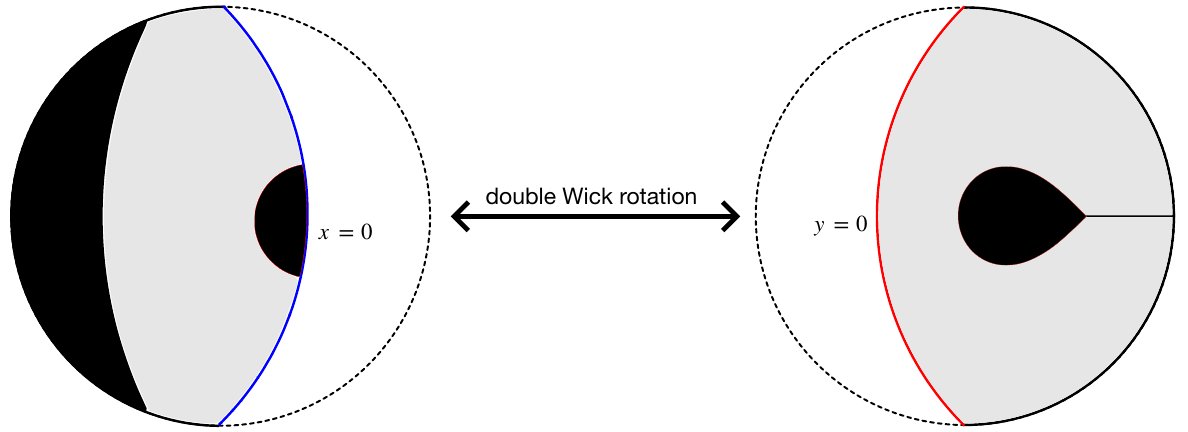}
    \caption{In most cases the structure of the bulk is as pictured in figure \ref{quSol1}, but in case I for the $x=0$ brane and case III for the $y=0$ brane the structure is different. Here we illustrate these two cases, which are related by the double analytic continuation.  In case I we are only drawing the region between the two horizons. See figure \ref{fig:xy_diagrams} for a more complete picture of the structure.  }
    \label{fig:DW_2horizon_to_string}
\end{figure}

The brane intersects the bulk black hole horizon at $y=y_-$, and we have a compact horizon ending on the brane in all cases:  A spatial slice of this horizon at constant $t$ forms a disc, as pictured in figure \ref{fig:DW_2horizon_to_string}.\footnote{Without the brane, by contrast, the horizon at $y=y_-$ is compact only in case III.} More generally, the surfaces of constant $t,y$ for $y <0$ form discs, with a boundary on the brane. In the region $y > 0$, the surface at fixed $t,y$ is still a disc, but now the boundary is on the asymptotic boundary of the spacetime. We have $y \leq x \leq x_+$, so there is a family of such discs shrinking down towards the axis at $x=x_+$. In case I there is another horizon at $y = y_+$, which is a disc with a boundary on the asymptotic boundary of the spacetime, that is, a non-compact acceleration horizon. Case I corresponds to $\kappa =1$ and ${\ell}/{\ell_3} < {2}/({3\sqrt 3 \mu}) $, so as we take the limit $\ell \to 0$ for fixed $\mu$ for positive-mass quantum BTZ, we will always enter case I. The structure of the spacetime is illustrated in the left picture in figure \ref{quSol1} for cases II, III and IV, and in the left picture in figure \ref{fig:DW_2horizon_to_string} in case I. 

In \cite{Emparan:2020znc} the bulk metric in the presence of the brane was written in terms of the $r$ coordinate in \eqref{branec}. This coordinate system does not cover the whole bulk, only the portion $y <0$. In particular, the additional horizon at $y=y_+$ in case I is not covered. 

\subsubsection{$y=0$ brane: quantum soliton}

Consider now inserting a brane at $y=0$. The extrinsic curvature is $K_{ab} = A \sqrt{\lambda} h_{ab}$, so we have
\begin{equation} \label{ll3y}
    \ell = \frac{1}{A \sqrt{\lambda}} \quad \Rightarrow \quad \ell_3 = \frac{1}{A}. 
\end{equation}
We see that it only makes sense to insert a brane at $y=0$ for $\lambda >0$; this is because we need $\lambda >0$ to have $H(0) <0$, so that the brane is timelike. Inserting a positive-tension brane will restrict us to $y <0$. The metric on the $y=0$ surface is 
\begin{equation}
    ds_3^2 = \frac{1}{A^2 x^2} \left[ - \lambda dt^2 + \frac{dx^2}{G(x)} + G(x) d\phi^2\right] .
\end{equation}
Setting
\begin{equation} \label{solbcoord}
    x = \frac{1}{A r}, \quad \phi = A \varphi, \quad t = \frac{\tau}{\sqrt{\lambda}}, \quad \quad \kappa = k, \quad   \hat \mu = \mu \sqrt{\lambda}, 
\end{equation}
we will recover the metric \eqref{qsoliton} on the brane in $\tau,r,\varphi$ coordinates, with $f(r) = {G(x)}/{x^2}$. This can also be understood from a double analytic continuation: the double analytic continuation of the C-metric (\ref{contxy},\ref{contpar}) interchanges $x$ and $y$, turning the $x=0$ brane into the $y=0$ brane up to a change of parameters \eqref{contpar}.  

On the $y=0$ brane we have $x \geq 0$, so $r \in (0, \infty)$ covers the whole brane, but again this coordinate does not cover the whole bulk. Since we are restricted to the region $y<0$, the only horizon in the spacetime with the brane is the one at $y=y_-$. This is a non-compact horizon with a boundary at infinity in cases I, II, and IV, and a compact black hole horizon in the bulk away from the brane in case III, with a conical singularity along the axis at $x=x_-$, extending from the horizon to the asymptotic boundary. The bulk geometry for cases I, II and IV is illustrated in figure \ref{quSol1}, while case III is illustrated in figure~\ref{fig:DW_2horizon_to_string}.


\subsubsection{Two brane scenarios}

\begin{figure}
    \centering
    \includegraphics[width=0.40\linewidth]{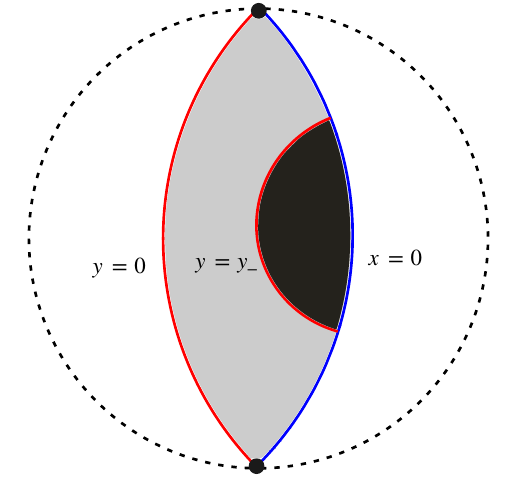}
    \includegraphics[width=0.50\linewidth]{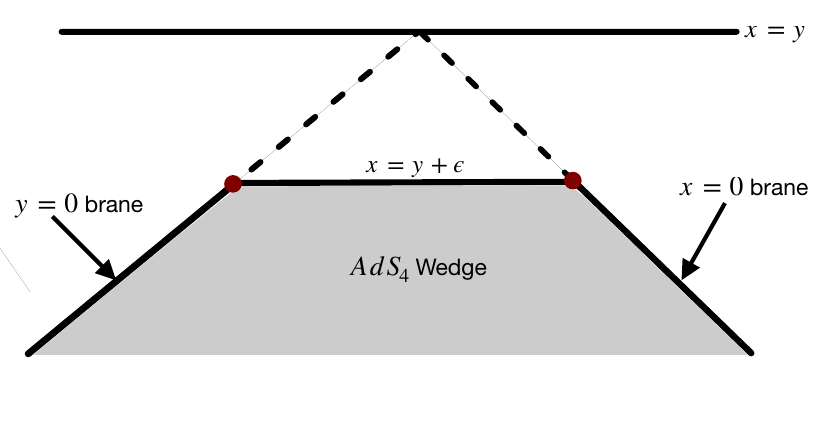}
    \caption{\textbf{Left}: Sketch of the wedge braneworld construction with both $x=0$ and $y=0$ branes. The bulk geometry is cut off at both brane locations and only the grey-shaded region is kept. \textbf{Right}: Zoomed-in view of the same construction. The conformal boundary lies at $x=y$, and we introduce a cutoff surface at $x=y+\epsilon$, which will be important in the action calculation in section \ref{sec:thermo}.}
    \label{QuantumBTZ}
\end{figure}

We can also consider inserting both a brane at constant $x$ and a brane at constant $y$. In \cite{Emparan:1999fd}, the case with an $x=0$ and a $y=0$ brane was considered. We are then restricted to $y<0$ and $x>0$ in the bulk. The bulk spacetime in all cases has a single horizon at $y=y_-$, which intersects the $x=0$ brane. The conformal boundary of the C-metric is almost entirely removed, only the ring at $x=y=0$ remains. The resulting spacetime is pictured in figure \ref{QuantumBTZ}.  A boundary description could be given in the framework of wedge holography \cite{Akal:2020wfl}, with the dual being a two-dimensional CFT living at $x=y=0$, but we will not explore this here.  

We will mostly focus on this scenario in the following, but other possibilities exist. If $\xi_c >0$, in cases I and II, and we insert branes at $x=0$ and $y=\xi_c$, we will retain a portion of the conformal boundary, at $0 \leq x=y \leq \xi_c$. If we consider instead inserting branes at $x=\xi_c$ and $y=0$, the two branes will intersect and there is no conformal boundary.  If $\xi_c <0$, in cases III and IV, we retain a portion of the conformal boundary when we insert branes at $x=\xi_c$ and $y=0$, and the branes intersect if we insert branes at $x=0$ and $y=\xi_c$. The cases with a portion of conformal boundary are intermediate in structure between the single brane scenario and the two-brane scenario with branes at $x=0$ and $y=0$. The cases with intersecting branes seem harder to interpret. 

\subsection{Zero backreaction limit}
\label{Schwlimit}

We can write the bulk C-metric in coordinates which are related to the Schwarzschild-AdS and AdS soliton geometries, which makes it easy to relate the brane C-metric constructions to the unbackreacted solutions in section \ref{SchwAdS}. The effective theory on the brane is a good description for ${\ell}/{\ell_3} \ll 1$, and we expect to recover the unbackreacted solutions in the limit as  ${\ell}/{\ell_3} \to 0$. In the brane theory, it is natural to think about the limit in units with $\ell_3$ fixed, and take $\ell \to 0$, as in \cite{Emparan:2020znc}. But in the bulk description, it is natural to describe the limits in units where $\ell, \ell_4$ are fixed and $\ell_3 \to \infty$; the AdS length scale on the brane diverges as the brane goes off to the boundary. 

We have two different branes, which have different values of $\ell$, so there are two different `unbackreacted' limits in the parameter space. For the $x=0$ branes, we see from \eqref{ll3x} that the limit ${\ell}/{\ell_3}  \to 0$ is the limit $\lambda \to 0$. This limit was discussed in \cite{Emparan:2020znc}. For the $y=0$ branes, we see from \eqref{ll3y} that ${\ell}/{\ell_3}  \to 0$ is the limit $\lambda \to \infty$. If we hold $\ell_4$ fixed, this is also the limit as $A \to 0$, so this is the zero acceleration limit of the C-metric, where it reduces to Schwarzschild-AdS.  

We now develop coordinates adapted to each of these limits. To make our discussion independent of the choice of units and to facilitate comparison to the discussion in section \ref{SchwAdS}, it is useful to take the brane AdS scale out as an overall factor. That is, for the $x=0$ brane, in \eqref{qBTZ} we write 
\begin{equation}
    r = \ell_3 \bar r, \quad  t = \ell_3 \bar t, \quad \phi =  \bar \phi, 
\end{equation}
 so
 \begin{equation}
     ds_3^2 = \ell_3^2 (- f(\bar r) d\bar t^2 + \frac{d\bar r^2}{f(\bar r)} + \bar r^2 d\bar \phi^2 ), 
 \end{equation}
with 
\begin{equation}
    f(\bar r) = \bar r^2 - \kappa - \frac{\ell \mu}{\ell_3 \bar r} = \bar r^2 +k - \frac{\sqrt{\lambda}\mu}{\bar r}. 
\end{equation}
In terms of the C-metric coordinates, we set 
\begin{equation}
    y = - \frac{1}{A \ell_3 \bar r} = - \frac{\sqrt{\lambda}}{\bar r}, \quad t = A\ell_3 \bar t = \frac{\bar t}{\sqrt{\lambda}}, \quad \phi = \bar \phi.
\end{equation}
The $x=0$ brane restricts us to $x>0$, and we can further define 
\begin{equation}
    x = \frac{\ell_4}{\rho} = \frac{1}{A \sqrt{\lambda+1} \rho} \quad \Rightarrow \quad \frac{G(x)}{x^2} = F(\rho),  
\end{equation}
with $F(\rho)$ given by \eqref{SchwF} with $\kappa = -k$ and $2G_4 M = \mu \ell_4$. With this choice of coordinates, the C-metric becomes 
\begin{equation}
    ds^2 = \Omega^2 \left[ F(\rho) \ell_4^2 d\bar \phi^2 + \frac{d\rho^2}{F(\rho)} + \frac{\rho^2}{\bar r^2} \left( - f(\bar r) d\bar t^2 + \frac{d\bar r^2}{f(\bar r)} \right) \right], 
\end{equation}
where 
\begin{equation}
    \Omega = \frac{ \sqrt{\lambda +1}}{1 + \frac{ \sqrt{\lambda} \rho}{\ell_4 \bar r}}. 
\end{equation}

This coordinate transformation rewrites the C-metric for small $\lambda$ as a deformed version of the AdS soliton. 
For $\lambda =0$, this is precisely the bulk AdS soliton \eqref{adssol} with  a portion of the boundary written in locally AdS$_3$ coordinates. As discussed in the previous section, this coordinate system only covers the region with $y <0$ in the bulk. In the limit as $\lambda \to 0$, this is precisely the half of the bulk covered by writing the boundary in BTZ coordinates. 

The effect of non-zero $\lambda$ is to deform the geometry through the additional term in $f(\bar r)$ and an overall conformal rescaling of the geometry. The $x=0$ brane in these coordinates is always at $\rho \to \infty$, but for non-zero $\lambda$ the effect of the overall conformal factor is to bring this in to finite position, turning the non-dynamical boundary for $\lambda=0$ into a dynamical brane. In the C-metric, $\phi$ is a periodic coordinate, with period \eqref{phiCper},  which is precisely the period \eqref{phiFper} which makes the AdS soliton \eqref{adssol} smooth at $\rho=\rho_+$.  

For $\kappa=1$, the unbackreacted geometry on the boundary is a BTZ black hole, which already had a horizon; the back-reaction for non-zero $\lambda$ is a quantitative but not qualitative modification. For $\kappa=-1$, the unbackreacted geometry was a conical defect, as discussed in section \ref{SchwAdS}. Here the back-reaction modifies the geometry more significantly, replacing the defect with a horizon. In the limit as $\lambda \to 0$, the horizon shrinks. The solution away from the horizon goes over smoothly to the conical defect spacetime, the temperature of the horizon diverges in this limit (as when we consider the zero mass limit of a Schwarzschild black hole in flat space), so the limiting procedure is somewhat more subtle than in the $\kappa=1$ case.

For the $y=0$ brane, we similarly rescale the coordinates in the quantum soliton geometry \eqref{qsoliton}, setting 
\begin{equation}
    r = \ell_3 \bar r, \quad  \tau = \bar \tau, \quad \varphi =  \ell_3 \bar \varphi, 
\end{equation}
 so
 \begin{equation}
     ds_3^2 = \ell_3^2 ( f(\bar r) d\bar \varphi^2 + \frac{d\bar r^2}{f(\bar r)} - \bar r^2 d\bar \tau^2 ), 
 \end{equation}
with 
\begin{equation}
    f(r) = \bar r^2 - \kappa - \frac{\ell \hat \mu}{\ell_3 \bar r} = \bar r^2 - k - \frac{ \hat \mu}{\sqrt{\lambda} \bar r} 
\end{equation}
where $\kappa = k$ and $\hat \mu = \mu \sqrt{\lambda}$. We then want to take the limit $\frac{\ell}{\ell_3} \to 0$ at fixed $\hat \mu$, not fixed $\mu$, in order to get the metric on the brane to go to the locally AdS$_3$ form. In terms of bulk coordinates, introducing this coordinate system on the $y=0$ brane corresponds to setting
\begin{equation} 
    x = \frac{1}{A \ell_3 \bar r} = \frac{1}{\bar r}, \quad \phi = A \ell_3 \bar \varphi= \bar \varphi, \quad t = \frac{\bar\tau}{\sqrt{\lambda}}.
\end{equation}
Since $y=0$ brane restricts us to $y<0$,  we further define 
\begin{equation}
    y = -\frac{\sqrt{\lambda} \ell_4}{\rho} = -\frac{\sqrt{\lambda}}{A \sqrt{\lambda+1} \rho} \quad \Rightarrow \quad -\frac{H(y)}{y^2} = F(\rho),  
\end{equation}
where now $F(\rho)$ is given by \eqref{SchwF} with $\kappa =k$ and $2G_4M = \mu \sqrt{\lambda} \ell_4 = \hat \mu \ell_4$. We thus see that holding $\hat \mu$ fixed as $\lambda \to \infty$ seems natural in the bulk in these coordinates, as it corresponds to holding $G_4 M/\ell_4$ fixed. With this choice of coordinates, the C-metric becomes 
\begin{equation}
    ds^2 = \Omega^2 \left[ -F(\rho) \ell_4^2 d\bar \tau^2 + \frac{d\rho^2}{F(\rho)} + \frac{\rho^2}{\bar r^2} \left( f(\bar r) d\bar \varphi^2 + \frac{d\bar r^2}{f(\bar r)} \right) \right], 
\end{equation}
where 
\begin{equation}
    \Omega = \frac{ \sqrt{\lambda +1}}{\sqrt{\lambda} + \frac{ \rho}{\ell_4 \bar r}}. 
\end{equation}

This coordinate transformation rewrites the C-metric for large $\lambda$ as a deformed version of Schwarzschild-AdS. In the limit as $\lambda \to \infty$ with $\hat \mu$ fixed, $\Omega \to 1$ and this reduces to the Schwarszchild-AdS metric \eqref{SAdS} with the boundary written in locally AdS$_3$ coordinates. The effect of finite $\lambda$ is to deform the geometry through the additional term in $f(\bar r)$ and an overall conformal rescaling of the geometry. The $y=0$ brane in these coordinates is always at $\rho \to \infty$, but for finite $\lambda$ the effect of the overall conformal factor is to bring this in to finite position, turning the non-dynamical boundary for $\lambda \to \infty$ into a dynamical brane. 

These two forms of the C-metric are related through the double analytic continuation (\ref{contxy},\ref{contpar}) which interchanges $x$ and $y$. 

As for the $x=0$ branes, the effect of the deformation at large $\lambda$ is more significant for $\kappa=-1$ than for $\kappa=1$. For $\kappa=1$ the unbackreacted geometry on the boundary is global AdS$_3$. At finite $\lambda$, the geometry on the brane still has a smooth origin, although the geometry is modified. For $\kappa=-1$, the unbackreacted geometry has a Rindler-like acceleration horizon, with the CFT at a mismatched temperature. At finite $\lambda$, this horizon is capped off, replaced with a smooth origin. As for the $x=0$ branes, there is a subtlety in the limiting procedure for $\kappa=-1$. Here the issue is that the period of the $\bar \varphi$ coordinate vanishes as we take the limit. For $k=-1$, $G(x_+) = 1+ x_+^2 - \mu x_+^3 =0$ implies that as $\mu \to 0$, $x_+ \approx 1/\mu$, and hence 
\begin{equation}
    \Delta \phi = \frac{4\pi}{3\mu x_+^2 - 2 x_+} \approx 4\pi \mu. 
\end{equation}
So the period of the $\bar \varphi$ coordinate in the limit as $\lambda \to \infty$ goes to zero like $1/\sqrt{\lambda}$ for fixed $\hat \mu$. This periodicity is related under the double analytic continuation to the inverse temperature of the black hole horizon in the $x=0$ case; as is usual the double analytic continuation converts a feature of the quantum state into a geometric property of the solution. In the metric \eqref{H2}, the angle $\varphi$ is not periodically identified, so the geometry for finite $\lambda$ only locally approaches the hyperbolic Schwarzschild-AdS one in the limit.


For the C-metric, the bulk geometry is only well-behaved for $\mu >0$. In the $x=0$ branes with $\kappa=-1$, this means that we have a backreacted version of the conical deficits, but not the conical excesses. For the $y=0$ branes with $\kappa=-1$, it means we have a backreacted version of the case where the temperature of the CFT state higher than the Rindler temperature of the horizon, but not lower. For the conical defects it seems physically reasonable that we would only have a solution for conical defects; the conical excesses have masses below that of global AdS$_3$, so it would be surprising to find regular backreacted geometries with these asymptotics. For the Rindler horizon, it is less clear that the state with lower temperature is less physical than that with higher temperature, but since it is related to the case of conical defects/excesses by analytic continuation, that may be the case.

\subsection{Geometry of the conformal boundary}
\label{Cboundary}

In the cases with a single brane (and in some two-brane cases), the spacetime also retains a portion of the conformal boundary.  In the effective brane theory, we then have the effective theory on the brane coupled to the holographic CFT on the fixed curved geometry on the remaining conformal boundary, with a transparent boundary condition between the CFT on the brane and the CFT on the boundary. It is therefore important to describe the geometry of the conformal boundary. In previous discussions of quantum BTZ, this issue has not been considered in detail. Here we point out that the geometry on the conformal boundary is non-trivial, and depends on the parameters of the bulk solution. In particular, the geometry of the conformal boundary at finite $\lambda$ differs from that in the no backreaction limit. So the deformation going to finite $\lambda$ is not just a question of making the geometry on half of the conformal boundary dynamical, turning it into a brane. Also the non-dynamical geometry on the other half of the conformal boundary gets modified. One of the advantages of the scenario with two branes at $x=0$ and $y=0$ is that it avoids this complication. 

Taking out $\frac{1}{A^2(x-y)^2}$ as a conformal factor and setting $x=y = \xi$, the metric on the conformal boundary is
\begin{equation} \label{cbxy}
    ds_3^2 = G(\xi) d\phi^2 - \frac{\lambda+1}{G(\xi) H(\xi)} d\xi^2 + H(\xi) dt^2. 
\end{equation}
This metric depends non-trivially on both $\lambda$ and $\mu$. In this section, we will focus on the dependence on $\lambda$, characterising how the general metric on the conformal boundary differs from the metric obtained in the limits $\lambda \to 0, \infty$. 

For the $x=0$ branes, we keep the portion of the conformal boundary at $\xi>0$. Consider the case with $k=-1$, corresponding to $\kappa=1$. Inspired by the relation to the AdS soliton for $\lambda=0$, introduce the coordinates
\begin{equation}
    \xi = \sqrt{\lambda} \cos \theta, \quad t = \frac{\bar t}{\sqrt{\lambda}},
\end{equation}
so $\xi >0$ corresponds to $\theta < \frac{\pi}{2}$. Then 
\begin{equation}
    H(\xi) = - \lambda (\sin^2 \theta + \mu \sqrt{\lambda}  \cos^3 \theta) = - \lambda \sin^2 \theta h(\theta), \quad h(\theta) = 1 + \mu \sqrt{\lambda}  \frac{\cos^3 \theta}{\sin^2 \theta} 
\end{equation}
and 
\begin{equation}
    G(\xi) = 1 + \lambda \cos^2 \theta - \mu \lambda^{3/2} \cos^3 \theta = g(\theta). 
\end{equation}
The metric on the conformal boundary in these coordinates is then 
\begin{equation}
    ds_3^2 = g(\theta)  d\phi^2 + \frac{\lambda+1}{g(\theta) h(\theta)} d\theta^2 - h(\theta) \sin^2 \theta d\bar t^2. 
\end{equation}
We see that if $\lambda=0$, this reduces to half of the dS$_2 \times S^1$ conformal boundary of the AdS soliton \eqref{adssol}, with $\theta \in (0, \pi/2)$. But for finite $\lambda$ it is deformed. In particular, while for $\lambda=0$ $t$ becomes null at $\theta=0$, for non-zero $\lambda$ this root disappears, and instead we have $\xi \in (0, x_+)$ in case II and $\xi \in (0, y_+)$ in case I. Typically in either case the whole range of $\xi$ is not covered by the $\theta$ coordinate introduced above; the range is extended. In case II the conformal boundary closes off smoothly at $\xi=x_+$, where the $\phi$ circle shrinks. This is a smooth origin as $H(x_+) = G(x_+) - (\lambda +1) = - (\lambda+1)$.  In case I, the conformal boundary has an event horizon at $\xi = y_+$, where the $t$ direction becomes null. As $\lambda \to 0$, $y_+ \approx \sqrt{\lambda} (1 + \frac{1}{2} \mu \lambda)$, and this goes smoothly over to the root at $\theta=0$, and we recover the dS$_2 \times S^1$ boundary of the AdS soliton. 

Consider the case $k=1$, corresponding to $\kappa=-1$. We then introduce coordinates 
\begin{equation}
    \xi = \sqrt{\lambda} \sinh \theta, \quad t = \frac{\bar{t}}{\sqrt{\lambda}}. 
\end{equation}
Then 
\begin{equation}
    H(\xi) = - \lambda (\cosh^2 \theta + \mu \sqrt{\lambda}  \sinh^3 \theta) = - \lambda \cosh^2 \theta h(\theta), \quad h(\theta) = 1 + \mu \sqrt{\lambda}  \frac{\sinh^3 \theta}{\cosh^2 \theta} 
\end{equation}
and 
\begin{equation}
    G(\xi) = 1 - \lambda \sinh^2 \theta - \mu \lambda^{3/2} \sinh^3 \theta = g(\theta). 
\end{equation}
The metric on the conformal boundary in these coordinates is then 
\begin{equation}
    ds_3^2 = g(\theta)  d\phi^2 + \frac{\lambda+1}{g(\theta) h(\theta)} d\theta^2 - h(\theta) \cosh^2 \theta d\bar t^2. 
\end{equation}
This again reduces to the boundary of \eqref{hypsol} when $\lambda=0$, where $\theta \in (0, \infty)$. For finite $\lambda$, the range of $\xi$ is restricted by the appearance of a root in $G(\xi)$, $\xi \in (0, x_+)$, which corresponds to a restriction on the range of $\theta$, $\theta \in (0, \sinh^{-1}(x_+/\sqrt{\lambda}))$. At $\xi=x_+$ the boundary closes off smoothly as the $\phi$ circle shrinks. 

For $y=0$ branes, we keep the portion of the boundary at $\xi<0$. This introduces some sign changes; also we are now interested in the limit $\lambda \to \infty$. For $k=1$, which for $y=0$ branes corresponds to $\kappa=1$, we set 
\begin{equation}
    \xi = - \cos \theta, \quad t = \frac{\tau}{\sqrt{\lambda}}. 
\end{equation}
Then 
\begin{equation}
    G(\xi) = \sin^2 \theta + \mu \cos^3 \theta = \sin^2 \theta + \frac{\hat \mu}{\sqrt{\lambda}} \cos^3 \theta = \sin^2 \theta g(\theta), \quad g(\theta) = 1 + \frac{\hat \mu}{\sqrt{\lambda}}  \frac{\cos^3 \theta}{\sin^2 \theta} 
\end{equation}
and
\begin{equation}
    H(\xi) = -\lambda \left[ 1 + \frac{\cos^2 \theta}{\lambda} - \frac{\hat \mu \cos^3 \theta}{\lambda^{3/2}}\right] = - \lambda h(\theta). 
\end{equation}
The metric on the conformal boundary in these coordinates is then 
\begin{equation}
    ds_3^2 = g(\theta) \sin^2 \theta  d\phi^2 + \frac{\lambda+1}{\lambda g(\theta) h(\theta)} d\theta^2 - h(\theta) d\tau^2. 
\end{equation}
We see that as $\lambda \to \infty$ at fixed $\hat \mu$, this reduces to half of the ESU conformal boundary in \eqref{ESU}, with $\theta \in (0, \pi/2)$. It has a smooth origin at $\theta=0$, corresponding to $\xi=x_- = -1$. For finite $\lambda$, this is no longer a root, and the range of $\xi$ is extended, and not fully covered by the $\theta$ coordinate. In case III, we still have a root of $G(\xi)$ at $\xi=x_- < -1$, and the boundary terminates there with the $\phi$ circle closing off. But this is now a conical defect, as the period of $\phi$ is chosen to make the bulk geometry smooth at $\xi=x_+$. In case IV, we no longer have a negative root for $G(\xi)$ and we instead reach a root of $H(\xi)$ at $\xi= y_-$, where the $\tau$ direction becomes null, so we have a horizon on the boundary. 

For $k=-1$, corresponding to $\kappa=-1$, we define 
\begin{equation}
    \xi = - \sinh \theta, \quad t = \frac{\tau}{\sqrt{\lambda}}. 
\end{equation}
Then 
\begin{equation}
    G(\xi) = \cosh^2 \theta + \hat{\mu}{\sqrt{\lambda}}  \sinh^3 \theta = \cosh^2 \theta g(\theta), \quad g(\theta) = 1 + \frac{\hat \mu}{\sqrt{\lambda}}\frac{\sinh^3 \theta}{\cosh^2 \theta} 
\end{equation}
and
\begin{equation}
    H(\xi) = - \lambda (1 - \frac{\sinh^2 \theta}{\lambda} - \frac{\hat \mu}{\lambda^{3/2}}   \sinh^3 \theta) = - \lambda  h(\theta).
\end{equation}
The metric on the conformal boundary in these coordinates is then 
\begin{equation}
    ds_3^2 = g(\theta) \cosh^2 \theta d\phi^2 + \frac{\lambda+1}{\lambda g(\theta) h(\theta)} d\theta^2 - h(\theta)  d\tau^2. 
\end{equation}
For $\lambda \to \infty$ at fixed $\hat \mu$, this reduces to half of the boundary of the $\kappa=-1$ black hole in the coordinates of \eqref{kminusb}. For $k=-1$ $G(\xi)$ never has a negative root, so for finite $\lambda$ this is modified by developing a horizon at $\xi = y_-$, where the $\tau$ direction becomes null. 

The metric on the conformal boundary in the single brane cases is qualitatively different from that in the no backreaction limit. It is not clear how much effect this modification of the geometry on the conformal boundary is having on the theory on the brane. If we consider the situation with both the $x=0$ and $y=0$ branes, this problem does not arise.  An interesting direction for future development would be to see if one can obtain solutions of the classical equations of motion in the bulk keeping the conformal boundary fixed as we vary $\lambda$. This would presumably require numerical work along the lines of \cite{Horowitz:2018coe}.

\section{Quantum soliton}
\label{sec:qsol}

Having understood how we obtain it from a bulk C-metric, in this section we summarise the physical structure of the quantum soliton solution \eqref{qsoliton}. 

\subsection{Back-reacted global AdS$_3$}

Consider first the case with $\kappa=1$, where the unbackreacted solution is global AdS$_3$,
\begin{equation}
    ds_3^2 = - \cosh^2 \gamma d\tau^2 + d \gamma^2 + \sinh^2 \gamma d\varphi^2 = - r^2 d\tau^2 + \frac{dr^2}{r^2 -1} + (r^2-1) d\varphi^2. 
\end{equation}
In section \ref{SchwAdS}, we wrote the Schwarzschild-AdS solution in a form where this global AdS$_3$ metric is half the boundary. The bulk black hole gives a CFT stress tensor on the boundary
\begin{equation} 
    \langle T_{\mu}^{\ \nu} \rangle = \frac{M\ell_4}{8\pi\cosh^3 \gamma} {\rm diag}\left(-2,1,1  \right) = \frac{M\ell_4}{8\pi r^3} {\rm diag}\left(-2,1,1  \right). 
\end{equation}
The parameter $M$ corresponds to a choice of quantum state on AdS$_3$. The form of this thermal stress tensor is fixed by the symmetries in the ESU form of the metric. Thinking of this Schwarzschild-AdS solution as the $\lambda \to \infty$ limit of the C-metric with fixed $\hat \mu$, $2 G_4 M = \hat \mu \ell_4$. By moving from the $\lambda \to \infty$ limit to large but finite $\lambda$ at fixed $\hat \mu$, we obtain a back-reacted solution, where the metric on the brane becomes (in units with $\ell_3=1$) 
\begin{equation}
    ds_3^2 = - r^2 d\tau^2 + \frac{dr^2}{f(r)} + f(r) d\varphi^2, \quad f(r) = r^2 -1 -\frac{\hat \mu}{\sqrt{\lambda} r}. 
\end{equation}
This describes the back-reaction of the CFT stress tensor on the geometry.\footnote{And also the effect of the deformation of the metric on the other half of the conformal boundary, as discussed in the previous section.} 

At large $r$, the leading effect of the back-reaction is not the subleading term in $f(r)$, but rather the change in period of $\varphi$. The bulk solution (or equivalently, smoothness of the brane metric at the origin where $g_{\varphi\varphi} \to 0$) fixes 
\begin{equation}
    \Delta_\varphi = \frac{1}{A} \Delta \phi = \frac{4\pi \ell_3 x_+}{(3 - x_+^2)} = 2\pi \ell_3 \Delta, \quad \Delta = \frac{2x_+}{(3 - x_+^2)}.
\end{equation}
In the limit as $\lambda \to \infty$ for fixed $\hat \mu$, $x_+ \to 1$, so $\Delta \to 1$, and $\varphi$ has period $2\pi$. For finite $\lambda$, the period changes, which changes the ADM mass of the solution.  To make the asymptotics manifest, we should rescale the coordinate $\varphi = \frac{\Delta_\varphi}{2\pi} \bar \varphi$ so that $\bar \varphi$ is $2\pi$ periodic, define a radial coordinate $\bar r =  \frac{\Delta\varphi}{2\pi} \sqrt{f(r)}$ and set $\tau = \frac{\Delta}{\ell_3} \bar \tau$ to write the asymptotic metric in a canonical form. Asymptotically, we can ignore the last term in $f(r)$, so 
\begin{equation}
    \bar r^2 \approx \Delta^2 (r^2 - \ell_3^2 ), 
\end{equation}
and the metric becomes asymptotically 
\begin{equation}
    ds^2 \approx -(\frac{\bar r^2}{\ell_3^2} + \Delta^2) d\bar \tau^2 + \frac{d\bar r^2}{\frac{\bar r^2}{\ell_3^2}+\Delta^2} + \bar r^2 d\bar \varphi^2.  
\end{equation}
 Thus, the mass of the soliton is $M_3 = -\Delta^2$, in units where global AdS has $M_3=-1$. This change in mass reflects the back-reaction of the quantum stress tensor on the geometry, measuring the total amount of energy in the fluid on the brane. This is a non-trivial function of the energy density in the fluid, which is proportional to $\hat \mu$. On general grounds, the back-reaction, and hence the mass, should be a function of two parameters: $ {\ell}/{\ell_3} = {1}/{\sqrt{\lambda}}$, controlling the strength of the back-reaction, and $\hat \mu$, which characterises the state of the CFT matter on the brane. In fact $x_+$, and hence $M_3$, is determined by a single combination $\mu = \frac{\hat \mu}{\sqrt{\lambda}} = \frac{\hat \mu \ell}{\ell_3}$. For $\mu =0$, $\Delta =1$, so $M_3 =-1$. As $\mu \to \infty$, $x_+ \approx \mu^{-1/3}$, so $\Delta \to 0$ and hence $M_3 \to 0$. The root $x_+$ is a monotonically decreasing function of $\mu$ and $\Delta$ is a monotonically increasing function of $x_+$, so $M_3$ is a monotonically increasing function of $\mu$. As we increase the energy in the CFT stress tensor, the asymptotic mass increases, with $M_3 \in (-1,0)$.

The quantum solitons thus asymptotically approach a conical defect spacetime, but they are smooth in the interior. The unbackreacted AdS$_3$ solution had a smooth origin at $r=1$. As we turn on back-reaction, this remains smooth, but moves to larger $r$ because of the subleading correction in $f(r)$. Thus, the back-reaction of the CFT matter cuts off the geometry a little further out -- something we will see in a more dramatic form in the $\kappa=-1$ case. 

We can obtain the subleading terms in the quantum stress tensor on the brane by double analytic continuation of the results for quantum BTZ in \cite{Emparan:2020znc}. They are given as a formal expansion in powers of $\ell$, 
\be 
\langle T_a ^b \rangle = \langle T_a ^b \rangle_0  + \ell^2 \langle T_a ^b \rangle_2 + \cdots \, .
\ee
The subleading contribution is not traceless, as a result of the scale introduced by the cutoff $\ell$. We will not give the explicit expressions here, which are just given by double analytic continuation of the results of \cite{Emparan:2020znc}.

\subsection{The disappearing horizon}

Consider now the case with $\kappa=-1$. Here the unbackreacted solution has a horizon, 
\begin{equation}
    ds_3^2 = -\sinh^2 \gamma d\tau^2 + d\gamma^2 + \cosh^2 \gamma d\varphi^2 = - r^2 d\tau^2 + \frac{dr^2}{r^2+1}+ (r^2+1) d\varphi^2. 
\end{equation}
In section \ref{SchwAdS}, we wrote the hyperbolic Schwarzschild-AdS solution in a form where half of the boundary has this geometry. The bulk black hole gives a CFT stress tensor on the boundary 
\begin{equation} 
    \langle T_{\mu}^{\ \nu} \rangle = \frac{M \ell_4}{8\pi\sinh^3 \gamma} {\rm diag}\left(-2,1,1  \right) = \frac{M \ell_4}{8\pi r^3} {\rm diag}\left(-2,1,1  \right). 
\end{equation}
The parameter $M$ again corresponds to a choice of quantum state on the boundary. The stress tensor for $M \neq 0$ is not regular on the horizon. This is a Rindler-like horizon on the boundary, and we have chosen a state that is thermal in the Rindler coordinates, but at a different temperature from that corresponding to the Rindler acceleration. For $M >0$, we can obtain this solution as the $\lambda \to \infty$ limit of a regular C-metric solution with fixed $\hat \mu$, and $2 G_4 M = \hat \mu \ell_4$. The C-metric at large but finite $\lambda$ gives a back-reacted solution
\begin{equation}
    ds_3^2 = - r^2 d\tau^2 + \frac{dr^2}{f(r)} + f(r) d\varphi^2, \quad f(r) = r^2 +1 -\frac{\hat \mu}{\sqrt{\lambda} r}. 
\end{equation}
The first notable feature here is that while the unbackreacted solution had a horizon at $r=0$, at which the CFT stress tensor diverged, the backreacted solution has no horizon; instead the geometry is cut off by a smooth origin at $r>0$, and the CFT stress tensor is regular everywhere on the backreacted solution. 

This is related by double analytic continuation to the story for the quantum BTZ black hole for $\kappa=-1$, where the unbackreacted solution has a conical singularity (the analytic continuation of the non-regular horizon here) which upon back-reaction is concealed behind a horizon (whose analytic continuation gives the origin here). This is the first instance we are aware of where we can understand the back-reaction of matter at the wrong temperature on a horizon, and it is very interesting that the effect is to cut off the solution with a smooth origin.  

The unbackreacted solution has no identification of $\varphi$, but we can consider imposing a fixed period of $\varphi$ if we choose. The C-metric at finite $\lambda$ has $\varphi$ periodically identified, with 
\begin{equation}
    \Delta \phi = 2 \pi \ell_3 \Delta,\quad \Delta = \frac{2 x_+}{3 + x_+^2}. 
\end{equation}
As in the $\kappa=1$ case, this periodicity determines the asymptotic ADM mass of the brane geometry, $M_3 = \Delta^2$.  This is again a function of the single parameter $\mu = \frac{\hat \mu}{\sqrt{\lambda}} = \frac{\hat \mu \ell}{\ell_3}$. The root $x_+$ is still a monotonically decreasing function of $\mu$, with $x_+ \sim \mu^{-1}$ for $\mu \to 0$ and $x_+ \sim \mu^{-1/3}$ for $\mu \to \infty$, so $x_+ \in (0, \infty)$. However $\Delta$ is now not a monotonic function of $x_+$; it has a turning point at $x_+=\sqrt{3}$. So we have $M_3 \to 0$ for $\mu \to 0, \infty$, and a maximum at $\mu =\frac{4}{3 \sqrt{3}}$, where $M_3 = \frac{1}{3}$.

As in the $\kappa=1$ case, we can read off subleading terms in the CFT stress tensor by double analytic continuation of the results of \cite{Emparan:2020znc}. 

\section{Action and thermodynamics}
\label{sec:thermo}

For quantum BTZ, \cite{Emparan:2020znc} obtained a first law, relating the generalised entropy on the brane to the three-dimensional ADM mass. For the quantum soliton, there is no horizon, but we would still expect to find a similar first law, relating the entropy of the thermal CFT gas on the brane to the ADM mass calculated above. We indeed find such a first law, but this will arise in the scenario with two branes, at $x=0$ and $y=0$. There is then a single bulk thermodynamics, which can be interpreted in terms of the effective theory on the $x=0$ and $y=0$ branes in appropriate limits. This reproduces the discussion of quantum BTZ in \cite{Emparan:2020znc} in the former case and gives us a first law for the quantum soliton in the latter. 

The need for two branes is most easily seen by considering the single brane setup with a $y=0$ brane. The bulk has a non-compact horizon except in case III, so the entropy from the bulk perspective is infinite. This is clearly not reproduced by considering just the CFT gas on the $y=0$ brane. The bulk horizon intersects the conformal boundary, and the divergence in the bulk entropy is identified with the entropy of this horizon in the conformal boundary; it is the divergent entropy from entanglement in the CFT state across the horizon in this fixed background. Thus, to carry out a thermodynamic analysis from the CFT point of view in the single-brane setup clearly requires us to include both the CFT on the brane and the CFT on the conformal boundary. For the single brane setup with an $x=0$ brane, the horizon at $y=y_-$ has a finite area, but in case I there is also a non-compact horizon at $y=y_+$ whose entropy again corresponds to entanglement entropy in the CFT on the conformal boundary. Even in the cases where there isn't an additional horizon, thermodynamics in the bulk should be identified with that of the coupled system including the CFT on the brane and the CFT on the conformal boundary. 

Furthermore, the thermodynamics in the single-brane cases is complicated by the fact that varying the parameters in the C-metric solution isn't just changing the state of the CFT, it also changes the geometry of the conformal boundary, as noted in section \ref{Cboundary}. That is, the C-metric does not satisfy the Dirichlet boundary conditions usually assumed in black hole thermodynamics~\cite{Papadimitriou:2005ii,Compere:2008us}, as briefly noted in~\cite{Anabalon:2018qfv}. The thermodynamics of the one-brane setup should therefore be viewed as \textit{constrained}: the mass and entropy cannot be cleanly separated from additional work terms associated with deformations of the boundary geometry. 

The two-brane scenario with branes at $x=0$ and $y=0$ avoids this problem, as it removes the conformal boundary, as shown in figure \ref{QuantumBTZ}. The bulk thermodynamics is then just related to the theory on the branes.

\begin{figure}
    \centering
    \includegraphics[width=0.50\linewidth]{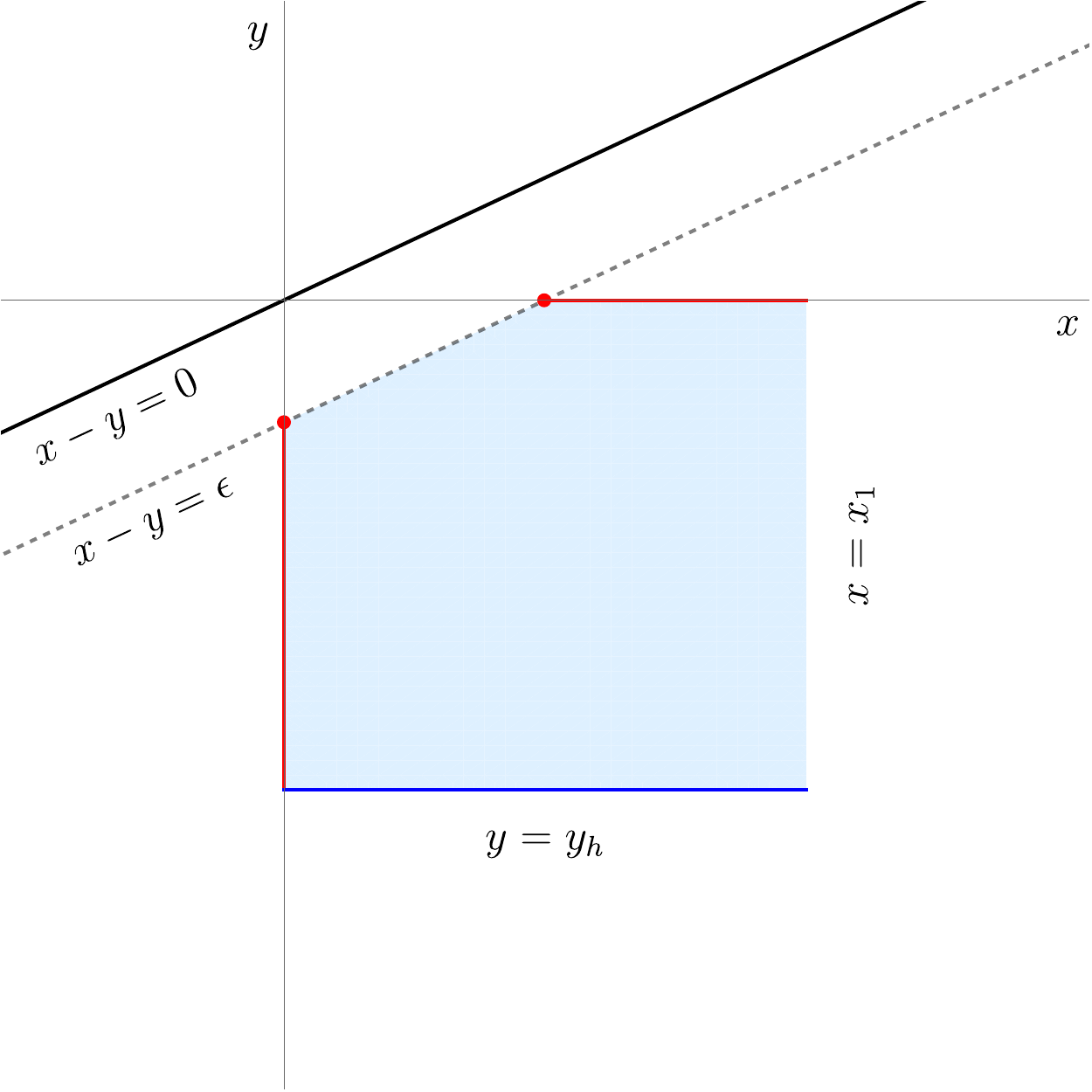}
    \caption{Region for the two-brane Euclidean action calculation in $(x,y)$ coordinates. The solid black line is the $z = 0$ asymptotic boundary, the dashed black line is the $z = \epsilon$ cutoff, the solid red lines are the branes, and the solid blue line is the Euclidean horizon. Solid red dots indicate the corners where the branes intersect the regulator surface.}
    \label{fig:two-brane-region}
\end{figure}


We will focus on the two-brane thermodynamics, leaving further consideration of the single-brane cases to future work. We therefore want to compute the bulk Euclidean action of the C-metric solution with branes at $x=0$ and $y=0$. The Euclidean C-metric is
\begin{equation} \label{xyCM1}
{\rm d} s^2 = \frac{1}{A^2 (x-y)^2} \left[-H(y) \dd \tau^2 - \frac{\dd y^2}{H(y)} + \frac{\dd x^2}{G(x)} + G(x) \dd \phi^2 \right] \, ,
\end{equation}
where 
\begin{equation} 
G(x) = 1 - k x^2 - \mu x^3 \, , \quad H(y) = - \left(\lambda + k y^2 + \mu y^3 \right) \, .
\end{equation}
Regularity of the Euclidean solution requires that $\tau \sim \tau + \Delta \tau$ and $\phi\sim \phi + \Delta \phi$ are periodic with
\be 
\Delta \tau = \frac{4 \pi}{|H'(y_-)|} \, , \quad \Delta \phi = \frac{4 \pi}{|G'(x_+)|} \, .
\ee
It will be useful to introduce two auxiliary parameters which simplify the expressions,
\be 
\nu \equiv \frac{\ell}{\ell_3} \,, \quad z \equiv - \frac{y_-}{\sqrt{\lambda} x_+}  \, .
\ee
Here, $y_-$ and $x_+$ are not independent but linked via the constraints $H(y_-) = 0$ and $G(x_+) = 0$. We use these constraints in simplifying the expressions below.

The region we wish to evaluate the Euclidean action on is shown in Figure~\ref{fig:two-brane-region}. We regulate the action by introducing a cutoff surface at \(z \equiv x-y=\epsilon\). Because this is a geometrically nontrivial region, a well-posed variational principle requires several additional terms: Gibbons--Hawking--York terms on the two branes and on the cutoff surface, corner terms at the intersections of the branes with the cutoff surface, and holographic renormalization counterterms on the cutoff surface. In addition, because the cutoff surface itself has boundaries where it meets the branes, one must also include the corresponding Gibbons--Hawking--York terms localized on those edges. The full action is therefore somewhat cumbersome, so for brevity we defer its explicit form and evaluation to Appendix~\ref{app:action}. The final result is
\begin{equation}
I_{E} = \frac{\Delta \tau \Delta \phi}{8\pi G_4 A^2}
\left[
\frac{\lambda+1}{(x_+-y_-)^2}
-\frac{1}{x_+^2}
-\frac{\lambda}{y_-^2}
\right] \, .
\end{equation}
This matches an earlier calculation by Kudoh and Kurita done via different methods~\cite{Kudoh:2004ub} --- see also~\cite{Panella:2024sor,Cao:2026jls}. This result has sometimes been interpreted to correspond to the action of a one-brane spacetime, but that interpretation is not correct. The action of a one-brane setup differs quantitatively from this. 

We want to describe the thermodynamics from the point of view of the branes. There will be two different descriptions, corresponding to the quantum soliton on the $y=0$ brane and the quantum BTZ black hole on the $x=0$ brane. Effective theories on these branes are valid for large and small $\lambda$ respectively, although the validity of the thermodynamics expressions below is not limited to these regimes. A key difference between the two is the normalization of the time coordinate, which effects the definition of the temperature and mass. For the quantum soliton ($y = 0$ brane), the transformations
\be 
t =  \frac{2 \pi \lambda}{\Delta \phi A} \tau \, , \quad \bar{r} = \frac{\Delta \phi}{2 \pi A x} \, , \quad \varphi = \frac{2 \pi}{\Delta \phi} \phi \, ,
\ee
bring the metric into the form
\be 
\dd s^2_3 = A^2\bar{r}^2 \dd t^2 + \frac{\dd \bar{r}^2}{f(\bar{r})} + \frac{f(\bar{r})}{A^2} \dd \varphi^2 
\ee
where
\be 
f(\bar{r}) = A^2 \bar{r}^2 -  \left(\frac{\Delta \phi}{2\pi}\right)^2 k - \left(\frac{\Delta \phi}{2\pi}\right)^3 \frac{\mu}{A \bar{r}} \, .
\ee
This is the canonical form of the metric, as $t$ reduces to the global time coordinate of AdS$_3$ asymptotically, and also exactly when $\mu = 0$ and $k = -1$. Tracking the transformations we have
\be \label{eqn:soliton_beta}
t \sim t + \beta \qquad \text{with} \quad \beta = \frac{2 \pi \sqrt{\lambda} \Delta \tau}{A \Delta \phi} \, .
\ee
Thus $T = 1/\beta$ is the temperature of the quantum fields residing on the quantum soliton geometry. Using that for the quantum soliton we identify $A = 1/\ell_3$ and writing the expression in terms of $\lambda = 1/\nu^2$ and $z$, we have
\be 
T_{\rm qSol} = \frac{1}{2 \pi \ell_3} \frac{z (2\nu + 3 z + z^3)}{\nu + 3 \nu z^2 + 2 z^3} \, .
\ee

\begin{figure}[t]
    \centering
    \includegraphics[width=0.49\linewidth]{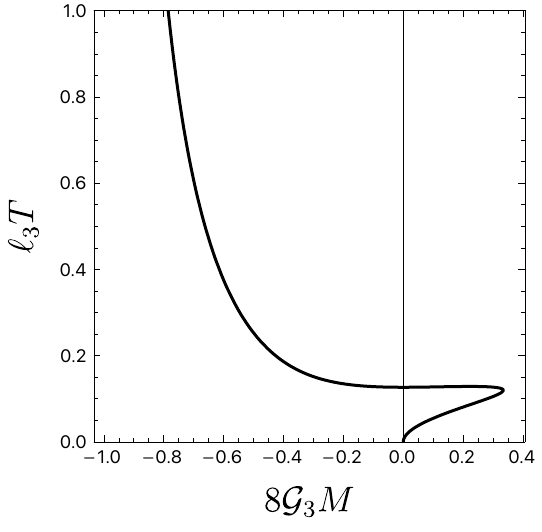}
    \includegraphics[width=0.49\linewidth]{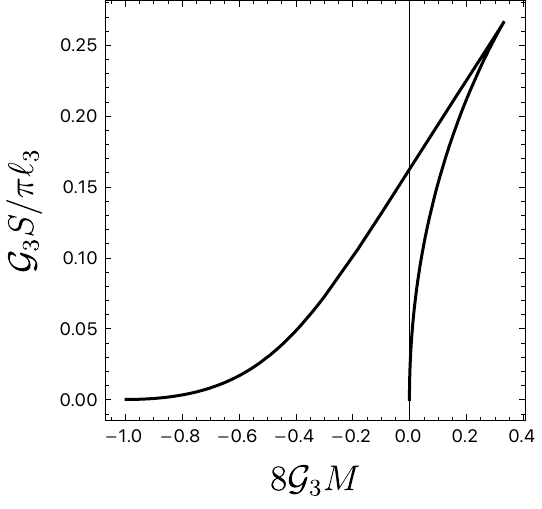}
    \caption{Left: A plot of the quantum soliton mass versus temperature for fixed $\nu = 2$. Those solitons with $M > 0$ correspond to $\kappa =+1$ while those solitons with $M < 0$ correspond to $\kappa = -1$. At fixed $\nu$, the $\kappa = +1$ branch approaches $M = 0$ as $T \to 0$, while the $\kappa = -1$ branch approaches $M = -1/(8 \mathcal{G}_3)$ as $T \to \infty$. As $\nu$ changes, the qualitative shape of the curve remains the same. Right: A plot of total entropy $S$ versus mass of the soliton for fixed $\nu=2$.
    }
    \label{fig:M_vs_T}
\end{figure}

Given a fixed inverse temperature $\beta$, the energy and entropy are computed from the on-shell action via the standard methods:
\begin{equation}
    M = \partial_\beta I_E \, , \quad S = \left(\beta \partial_\beta - 1\right)I_E \, .
\end{equation}
Applying this to the quantum soliton, and varying the action at fixed $A, \lambda$, we obtain
\be \label{eqn:soliton_mass}
 M_{\rm qSol} =  \frac{1}{2 \mathcal{G}_3} \frac{ z^2 (z + \nu)(\nu - z^3)}{(\nu + 3 \nu z^2 + 2 z^3)^2} \, , \qquad S_{\rm qSol} = \frac{\pi \ell_3}{\mathcal{G}_3} \frac{\nu z}{\nu + 3 \nu z^2 + 2 z^3} \, ,
\ee
where $\mathcal{G}_3$ was defined in~\eqref{eqn:calG3}. These quantities can be easily verified to satisfy the first law $dM = T dS$.  We plot the mass and entropy as functions of temperature in Figure~\ref{fig:M_vs_T}. 

In Section~\ref{sec:qsol}, we identified the mass of the quantum soliton from the asymptotic conical deficit on the brane, $8 \mathcal{G}_3 M = -\kappa \Delta^2$.
The mass obtained above from the Euclidean action agrees exactly with that result. Moreover, the entropy is precisely twice the area of the \(y<0\) portion of the bulk black hole horizon.  

We can alternatively describe the thermodynamics in terms of the $x=0$ brane. The metric on the $x = 0$ brane is brought into canonical form by the transformations
\be 
t =  \frac{2 \pi }{\Delta \phi  A} \tau \, , \quad \bar{r} = -\frac{\Delta \phi}{2 \pi A x} \, , \quad \varphi = \frac{2 \pi}{\Delta \phi} \phi \, ,
\ee
bringing the (Euclidean) metric into the form
\be 
\dd s^2_3 = f(\bar{r}) \dd t^2 + \frac{\dd \bar{r}^2}{f(\bar{r})} + \bar{r}^2 \dd \varphi^2 
\ee
where
\be 
f(\bar{r}) = A^2 \lambda  \bar{r}^2 +  \left(\frac{\Delta \phi}{2\pi}\right)^2 k - \left(\frac{\Delta \phi}{2\pi}\right)^3 \frac{\mu}{A \bar{r}} \, .
\ee
The correct temperature for the $x = 0$ brane is then
\be \label{eqn:soliton_beta}
t \sim t + \beta \qquad \text{with} \quad \beta = \frac{2 \pi \Delta \tau}{A \Delta \phi} \, .
\ee
Here this is interpreted as the temperature of the black hole horizon on the brane. Writing this explicitly and recalling that $A = 1/\ell$ and $\lambda = \nu$ for the quantum black hole we have 
\be 
T_{\rm qBTZ} = \frac{1}{2 \pi \ell_3} \frac{z (2 + 3 \nu z + \nu z^3)}{1 + 3  z^2 + 2 \nu z^3} \, .
\ee
Then the standard action computation yields the thermodynamic quantities
\begin{equation}
    M_{\rm qBTZ} = \frac{1}{2\mathcal{G}_{3}} \frac{z^{2}\bigl(1+\nu z\bigr)\bigl(1-\nu z^{3}\bigr)}
    {\bigl(1+3z^{2}+2\nu z^{3}\bigr)^{2}} \, ,
    \qquad
    S_{\rm qBTZ} = \frac{\pi \ell_3}{\mathcal{G}_{3}} \frac{z}
    {\bigl(1+3z^{2}+2\nu z^{3}\bigr)} \, .
\end{equation}
which satisfy the first law. These results agree with~\cite{Emparan:2020znc}. 

We note that the expressions for the mass, entropy, and temperature of the quantum soliton map formally to those of the quantum BTZ black hole under the transformation $\nu \to 1/\nu$. This follows from our earlier observation that the double Wick rotation of the C-metric exchanges the $x=0$ and $y=0$ branes, together with certain parameter rescalings. This, in turn, highlights a weak/strong back-reaction duality between the quantum soliton and the quantum BTZ black hole. 

Restricting to the brane, the asymptotic AdS$_3$ boundary geometries of the two solutions are the same: both are given by time times a circle. Under the conventional interpretation of the thermodynamics of these braneworld objects---see, for example, \cite{Emparan:2020znc,  Frassino:2022zaz, Frassino:2023wpc, HosseiniMansoori:2024bfi, Bhattacharya:2025tdn, Cao:2026jls}---the two solutions would therefore compete in the canonical ensemble.  Existing studies of quantum black hole thermodynamics typically consider phase transitions between the black hole and a thermal AdS$_3$ background. In most cases, however, the thermal AdS$_3$ solution is taken to be undeformed global AdS$_3$, having $M = -1/(8 \mathcal{G}_3)$. That background is then interpreted as describing the same quantum fields in a thermal state on AdS. But in that state the expectation value of the CFT stress tensor and the entropy both vanish. This seems unnatural: If there are sufficiently many quantum fields for their back-reaction to deform the black hole geometry significantly, then one should also expect those same thermal fields to back-react nontrivially on the vacuum geometry. This is precisely what the quantum soliton with $\kappa = +1$ represents: a smooth deformation of AdS$_3$ that incorporates the back-reaction of a large number of thermal quantum fields. In this sense, the soliton is a more natural candidate for the thermal AdS phase in this setting. Of course, as discussed in Section~\ref{Cboundary}, it is also likely that additional solutions with the same boundary conditions exist and could compete in the same ensemble, although constructing them will probably require numerical methods. We leave a more detailed investigation of the phase structure and possible transitions between these solutions for future work.

\section{Discussion}
\label{sec:disc}

In this paper, we have introduced a new semiclassical braneworld solution, the quantum soliton, obtained by placing a constant-tension brane on the $y=0$ surface of the AdS C-metric. Although its geometry is related to the quantum BTZ solution by double analytic continuation, its physical interpretation is significantly different. The quantum soliton provides a new exact holographic realization of CFT backreaction on asymptotically AdS$_3$ spacetimes, extending the braneworld programme beyond the black hole sector.

There are two different cases depending on the sign of $\kappa$. For $\kappa=1$, the zero-backreaction limit is global AdS$_3$, and the quantum soliton describes the smooth backreaction of a thermal CFT state on that geometry. In this sense, it provides a more natural realization of a thermal AdS phase in semiclassical gravity: once the CFT contains enough degrees of freedom to significantly affect the black-hole geometry, it is also natural to expect nontrivial backreaction on the vacuum background. The $\kappa=1$ soliton captures precisely this effect.

For $\kappa=-1$, the behaviour is even more striking. In the unbackreacted limit, the boundary geometry contains a Rindler-like horizon, and the corresponding CFT stress tensor is singular there. At finite backreaction, however, this horizon disappears and is replaced by a smooth origin. Thus, the quantum soliton furnishes a concrete example in which quantum backreaction resolves a would-be horizon singularity by capping off the geometry smoothly. This is a genuinely non-perturbative effect, and one of the most interesting features of the solutions we have constructed.

We also studied the Euclidean action and thermodynamics of the quantum soliton. We argued that the thermodynamics for both the quantum soliton and quantum BTZ are most naturally discussed in a two-brane scenario with branes at $x=0$ and $y=0$. We carefully calculated the Euclidean action for this two-brane scenario, showing that it leads to a first law for both branes. We argued that the $\kappa=1$ quantum soliton is the appropriate backreacted analogue of thermal AdS$_3$. The phase structure including this and quantum BTZ is an interesting direction for further study.

Along the way we have discussed a number of interesting features of these solutions. In the C-metric, we carefully discussed the possible root structures, observing the existence of an additional horizon in the bulk in case I, which was missed in previous brane discussions. We found new coordinate systems for the C-metric, adapted to the no backreaction limits for the two types of brane. We observed that there are subtleties in the zero backreaction limit for the $\kappa=-1$ solutions: for the $x=0$ brane the temperature of the black hole diverges as its size goes to zero, and for the $y=0$ brane the periodicity of the angular direction goes to zero as the backreaction goes to zero. We interpret this behaviour as a signal that in these cases, where the stress tensor is divergent on the unbackreacted solutions, it doesn't really make sense to think about turning the backreaction off: sufficiently close to the conical singularity or horizon, the effect of the backreaction from the CFT stress tensor is significant however small the gravitational coupling to the metric is. We considered the geometry of the conformal boundary, and noted that it depends on the bulk parameters, so it is not possible to study the thermodynamics in a conventional Dirichlet boundary condition. 

There are several directions in which this work could be extended. We can consider similar quantum solitons in the charged and rotating C-metrics. It would be interesting to explore the phase structure involving the quantum BTZ and quantum soliton solutions, and to determine whether additional saddles with the same asymptotic boundary conditions exist. It would also be useful to develop a clearer formulation of the one-brane action principle and its associated thermodynamics.  In the C-metric solutions one could consider relaxing the Dirichlet boundary conditions. It may then be possible to identify extra work terms associated with changing boundary conditions and recover a consistent thermodynamic description. It would be interesting to understand the relation to the work of \cite{Anabalon:2018qfv} from this perspective. Alternatively, one could seek to construct numerical solutions which satisfy Dirichlet boundary conditions, keeping the geometry of the conformal boundary fixed as the mass and temperature vary. Finally, it would be valuable to investigate whether similar quantum-soliton constructions exist in higher dimensions, and whether the disappearance of the $\kappa=-1$ horizon reflects a more general mechanism for the non-perturbative smoothing of singular semiclassical states.

\section*{Acknowledgements}

We thank Roberto Emparan, Antonia Frassino, Juan Pedraza, Jorge Rocha, Andrea Sanna, and Andrew Svesko for helpful discussions. RAH is supported by a Willmore Fellowship at Durham University. AKP is supported by the European Union’s Horizon Europe research and
innovation programme under the Marie Sk\l{}odowska-Curie grant agreement No.~101210745. The work of SFR is supported in part by STFC through grant number ST/X000591/1.

\appendix

\section{Euclidean action for the two-brane setup}
\label{app:action}

The Euclidean C-metric reads
\begin{equation} \label{xyCM}
{\rm d} s^2 = \frac{1}{A^2 (x-y)^2} \left[-H(y) \dd \tau^2 - \frac{\dd y^2}{H(y)} + \frac{\dd x^2}{G(x)} + G(x) \dd \phi^2 \right] \, ,
\end{equation}
where
\begin{equation}
G(x) = 1 - k x^2 - \mu x^3 \, , \qquad H(y) = - \left(\lambda + k y^2 + \mu y^3 \right) \, .
\end{equation}
We regulate the asymptotic divergence by placing a cutoff surface at $z \equiv x-y = \epsilon$, and we place branes at both $x=0$ and $y=0$. The region we consider is shown schematically in Figure~\ref{fig:two-brane-region}. The complete Euclidean action for this region reads (we include an overall factor of 2 because the branes are two-sided)
\begin{align}
    I_{\rm E} &= - \frac{1}{8 \pi G_4} \bigg[ \int_{\mathcal{M}} {\rm d}^4 x \sqrt{g} \left(\frac{6}{\ell_4^2} + R \right) + 2 \int_{x = 0} {\rm d}^3 x \sqrt{h_x} K_x + 2 \int_{y = 0} {\rm d}^3 x \sqrt{h_y} K_y 
    \nonumber
    \\
    &\qquad + 2 \int_{z = \epsilon} {\rm d}^3 x \sqrt{h_z} K_z
    + 2 \int_{J_{yz}} {\rm d}^2 x \sqrt{\sigma_{yz}} \left(\eta_{yz}-\eta_{yz}^{(0)} \right) + 2 \int_{J_{xz}} {\rm d}^2 x \sqrt{\sigma_{xz}} \left(\eta_{xz} -\eta_{xz}^{(0)}\right)
    \nonumber
    \\
    &\qquad - 2\int_{z = \epsilon} {\rm d}^3 x \sqrt{h_z} \left(\frac{2}{\ell_4} + \frac{\ell_4}{2} \mathcal{R}[h_z] \right)
    - 2 \ell_4 \int_{\partial(z=\epsilon)} {\rm d}^2 x \sqrt{\sigma}\,\mathcal{K} \bigg]
    \nonumber
    \\
    &\qquad + \int_{x = 0} {\rm d}^3 x  \sqrt{h_x} \tau_x + \int_{y = 0} {\rm d}^3 x  \sqrt{h_y} \tau_y \, .
\end{align}
Let us explain the terms in the action in the order they appear. Inside the brackets, the first term is the Euclidean Einstein-Hilbert action. The second, third, and forth terms are the Gibbons-Hawking-York contributions along the $x = 0$ brane, the $y=0$ brane, the the $z = \epsilon$ cutoff surface. The fifth and sixth terms are Hayward joint terms at the intersections of the $y = 0$ and $z = \epsilon$ surfaces which we denote $J_{yz}$ and the $x = 0$ and $z = \epsilon$ surfaces which we denote $J_{xz}$. The quantity $\eta$ is the angle between the surfaces, which we will define more carefully below, while $\eta^{(0)}$ is a counterterm at the joint. The seventh contribution is the counterterm on the $z = \epsilon$ cutoff surface.  The final term in the square brackets is the Gibbons--Hawking--York edge term associated with the curvature counterterm on the regulated surface $z=\epsilon$ --- here $\partial(z=\epsilon)=J_{xz}\cup J_{yz}$.  Since the counterterm contains a three-dimensional Einstein--Hilbert term for $h_z$ and the regulated hypersurface has boundaries at the joints, this edge term must be included for a well-posed Dirichlet variational problem. The final two contributions are the tension contributions of the $x = 0$ and $y = 0$ branes. Let us consider the computation of each of these terms in detail. Our convention for the extrinsic curvature is that it is computed with respect to the outward-pointing normal vector and $K_{ab} = h_a{}^c h_b{}^d \nabla_c n_d$.

\subsubsection*{Geometrical details}

Here we collect a few results that will be needed in several places below. We denote the outward-pointing unit normal one-forms to the $x = 0$, $y = 0$, and $z = \epsilon$ surfaces as $n_{(x)}, n_{(y)}$ and $n_{(z)}$, respectively. These are
\begin{align}
    n_{(x)} &= - \frac{{\rm d}x}{A (x-y) \sqrt{G(x)}} \, ,
    \\
    n_{(y)} &= \frac{{\rm d} y}{A (x-y) \sqrt{-H(y)}} \, ,
    \\
    n_{(z)} &= \frac{{\rm d}y - {\rm d}x}{A (x-y) \sqrt{G(x) - H(y)}} = -\frac{{\rm d} z}{A z \sqrt{G(x) - H(x - z)}} \, .
\end{align}
We also have the following determinants of the induced metrics on those slices:
\begin{align}
    \sqrt{h_x} &= \frac{\sqrt{G(0)}}{A^3 (-y)^3} \, ,
    \quad
    \sqrt{h_y} = \frac{\sqrt{-H(0)}}{A^3 x^3}  \, ,
    \quad
    \sqrt{h_z} = \frac{\sqrt{G(x) - H(x-\epsilon)}}{A^3 \epsilon^3} \, .
\end{align}

\subsection*{Bulk action}

To compute the bulk action, it is helpful to note that the Einstein equations enforce $R = 4 \Lambda$ which reduces the integral contribution to
\begin{equation}
\int_{\mathcal{M}} {\rm d}^4 x \sqrt{g} \left(\frac{6}{\ell_4^2} + R \right) = -\frac{6}{\ell_4^2} \int_{\mathcal{M}} {\rm d}^4 x \sqrt{g} = - \frac{6 \Delta \tau \Delta \phi}{\ell_4^2} \int {\rm d} x {\rm d} y \sqrt{g} \, .
\end{equation}
The metric determinant is $\sqrt{g}=1/[A^4(x-y)^4]$. We evaluate the integral by splitting it into two pieces:
\begin{align}
    \int_{\mathcal{M}} \frac{{\rm d} x {\rm d} y}{(x-y)^4} &= \int_{x = 0}^{x = \epsilon} {\rm d} x \int_{y = y_-}^{y = x-\epsilon} {\rm d}y \frac{1}{(x-y)^4} + \int_{x = \epsilon}^{x = x_+} {\rm d} x \int_{y = y_-}^{y = 0} {\rm d}y \frac{1}{(x-y)^4}
    \\
    &= \frac{1}{2 \epsilon^2} + \frac{1}{6 (x_+ - y_-)^2} - \frac{1}{6 x_+^2} - \frac{1}{6 y_-^2} \, .
\end{align}
Then we have for the bulk action
\begin{align}
\int_{\mathcal{M}} {\rm d}^4 x \sqrt{g} \left(\frac{6}{\ell_4^2} + R \right) &= - \frac{6 \Delta \tau \Delta \phi}{\ell_4^2 A^4} \left[\frac{1}{2 \epsilon^2} + \frac{1}{6 (x_+ - y_-)^2} - \frac{1}{6 x_+^2} - \frac{1}{6 y_-^2} \right]
\\
&= - \frac{6 \Delta \tau \Delta \phi(1 + \lambda)}{ A^2} \left[\frac{1}{2 \epsilon^2} + \frac{1}{6 (x_+ - y_-)^2} - \frac{1}{6 x_+^2} - \frac{1}{6 y_-^2} \right] \, .
\end{align}
In the second line we used
\begin{equation}
    \ell_4 = \frac{1}{A\sqrt{\lambda+1}} \, .
\end{equation}

\subsection*{GHY contribution on the $x = 0$ brane}

We compute the trace of the extrinsic curvature to be
\begin{equation}
    K_x = \frac{A \left( 6 G(0) - (x-y) G'(0) \right)}{2 \sqrt{G(0)}}  \quad \Rightarrow \quad \sqrt{h_x} K_x = \frac{\left( 6 G(0) + y G'(0) \right)}{2 A^2 (-y)^3} \, .
\end{equation}
Given that $G(0) = 1$ and $G'(0) = 0$, we then have the following integral for the GHY term along the $x = 0$ brane:
\begin{equation}
\int_{x= 0} {\rm d}^3 x \sqrt{h_x} K_x = \frac{3 \Delta \tau \Delta \phi}{A^2} \int_{y = y_-}^{y = - \epsilon} \frac{{\rm d} y}{(-y)^3} = \frac{3 \Delta \tau \Delta \phi}{2A^2} \left[\frac{1}{\epsilon^2}-\frac{1}{y_-^2} \right] \, .
\end{equation}

\subsection*{GHY contribution on the $y = 0$ brane}

We compute the trace of the extrinsic curvature to be
\begin{equation}
    K_y = -\frac{A \left(6 H(0) + x H'(0) \right)}{2 \sqrt{-H(0)}} \quad \Rightarrow \quad \sqrt{h_y} K_y = \frac{- \left(6 H(0) + x H'(0) \right)}{2 A^2 x^3} \, .
\end{equation}
Given that $H(0) = -\lambda$ and $H'(0) = 0$, we then have the following integral for the GHY term along the $y = 0$ brane:
\begin{equation}
\int_{y= 0} {\rm d}^3 x \sqrt{h_y} K_y = \frac{3 \lambda \Delta \tau \Delta \phi}{A^2} \int_{x = \epsilon}^{x = x_+} \frac{{\rm d} x}{x^3} = \frac{3 \lambda \Delta \tau \Delta \phi}{2A^2} \left[\frac{1}{\epsilon^2}-\frac{1}{x_+^2} \right] \, .
\end{equation}

\subsection*{GHY contribution on the $z = \epsilon$ cutoff}

The extrinsic curvature for a surface of constant $z$ reads
\begin{equation}
K_z
=3A\sqrt{G-H}-\frac{Az}{2(G-H)^{3/2}}
\Big((2G-H)\,H'(x-z)+(G-2H)\,G'(x)\Big),
\end{equation}
where each instance of $G$ is to be evaluated at $x$ and each instance of $H$ is to be evaluated at $(x-z)$. We then have
\begin{equation}
\sqrt{h_z}\,K_z
=\frac{3\,(G-H)}{A^2 z^3}
-\frac{1}{2A^2 z^2\,(G-H)}
\Big((2G-H)\,H'(x-z)+(G-2H)\,G'(x)\Big) \, .
\end{equation}
The resulting integral is somewhat complicated. We split it directly into divergent and finite parts and discard those terms that vanish in the $\epsilon \to 0$ limit. We have
\begin{equation}
    \int_{z= \epsilon} {\rm d}^3 x \sqrt{h_z} K_z = \frac{\Delta \tau\,\Delta \phi}{A^{2}}
\left[
\frac{3(\lambda+1)}{\epsilon^{2}}
+\frac{k(\lambda-1)}{2(\lambda+1)}
+\mathcal{O}(\epsilon^{2})
\right] \, .
\end{equation}

\subsection*{Holographic counterterms on $z = \epsilon$}

We have the standard holographic renormalization counterterm on the $z = \epsilon$ surface,
\begin{equation}
\int_{z = \epsilon} {\rm d}^3 x \sqrt{h_z} \left(\frac{2}{\ell_4} + \frac{\ell_4}{2} \mathcal{R}[h_z] \right) \, .
\end{equation}
The intrinsic curvature of that surface reads
\begin{align}
\mathcal{R}[h_z] &=
\frac{A^{2}\epsilon^{2}}{2\big(G(x)-H(x-\epsilon)\big)^{2}}
\Big[
-H(x-\epsilon)\,\big(G'(x)\big)^{2}
+G(x)\,\big(H'(x-\epsilon)\big)^{2}
\nonumber
\\
&\qquad
+2\big(G(x)-H(x-\epsilon)\big)\Big(G(x)\,H''(x-\epsilon)+H(x-\epsilon)\,G''(x)+G'(x)\,H'(x-\epsilon)\Big)
\Big] \, .
\end{align}
The evaluation of the integral reduces to a single integral over $x$ on the interval $[0,\epsilon]$. The result is messy, so we present only the divergent and finite parts in the $\epsilon \to 0$ limit:
\begin{equation}
    \int_{z = \epsilon} {\rm d}^3 x \sqrt{h_z} \left(\frac{2}{\ell_4} + \frac{\ell_4}{2} \mathcal{R}[h_z] \right)  = \frac{\Delta \tau\,\Delta \phi}{A^{2}}
\left[\frac{2 (\lambda+1)}{\epsilon^2} + \frac{k (\lambda-1)}{(\lambda+1)} + \mathcal{O}(\epsilon) \right] \, .
\end{equation}
In obtaining this form, we used $\ell_4 = 1/(A\sqrt{\lambda+1})$.

Because the counterterm contains a three-dimensional Einstein--Hilbert term for the induced metric $h_z$, and because the regulated surface $z=\epsilon$ ends on the branes at $J_{xz}$ and $J_{yz}$, we must also include the associated three-dimensional GHY term on the boundary of the regulated hypersurface:
\begin{equation}
\ell_4 \int_{\partial(z=\epsilon)}\dd^2x\,\sqrt{\sigma}\,\mathcal{K}
=
\ell_4 \left(
\int_{J_{xz}}\dd^2x\,\sqrt{\sigma_{xz}}\,\mathcal{K}_{xz}
+
\int_{J_{yz}}\dd^2x\,\sqrt{\sigma_{yz}}\,\mathcal{K}_{yz}
\right) \, ,
\end{equation}
where $\mathcal{K}$ is the trace of the extrinsic curvature of the edge, computed with the outward-pointing normal \emph{within} the $z=\epsilon$ hypersurface.

Using $y=x-\epsilon$ on $z=\epsilon$ and the induced two-metrics at the joints,
\begin{equation}
\sqrt{\sigma_{xz}}=\frac{\sqrt{-H(-\epsilon)}}{A^2\epsilon^2}\,,
\qquad
\sqrt{\sigma_{yz}}=\frac{\sqrt{\lambda G(\epsilon)}}{A^2\epsilon^2}\,,
\end{equation}
one finds the compact expressions
\begin{equation}
\sqrt{\sigma_{xz}}\,\mathcal{K}_{xz}
=\frac{H'(-\epsilon)}{2A\,\epsilon\,\sqrt{1-H(-\epsilon)}} \, ,
\qquad
\sqrt{\sigma_{yz}}\,\mathcal{K}_{yz}
=\frac{\lambda\,G'(\epsilon)}{2A\,\epsilon\,\sqrt{\lambda+G(\epsilon)}} \, .
\end{equation}
Expanding for small $\epsilon$ and keeping terms that survive as $\epsilon\to 0$ gives
\begin{equation}
\int_{J_{xz}}\dd^2x\,\sqrt{\sigma_{xz}}\,\mathcal{K}_{xz}
+
\int_{J_{yz}}\dd^2x\,\sqrt{\sigma_{yz}}\,\mathcal{K}_{yz}
=
\Delta \tau\Delta \phi\left[\frac{k(1-\lambda)}{A\sqrt{\lambda+1}}+\mathcal{O}(\epsilon)\right] \, .
\end{equation}
Using $\ell_4=1/(A\sqrt{\lambda+1})$, the edge contribution becomes
\begin{equation}
\ell_4 \int_{\partial(z=\epsilon)}\dd^2x\,\sqrt{\sigma}\,\mathcal{K}
=
\frac{\Delta \tau\Delta \phi}{A^2}
\left[\frac{k(1-\lambda)}{\lambda+1}+\mathcal{O}(\epsilon)\right] \, .
\end{equation}

\subsection*{Tension terms on the $x = 0$ and $y = 0$ branes}

The branes satisfy the Israel junction condition,
\begin{equation}
\left[K_{ab}\right] - h_{ab} \left[K\right] = 8 \pi G_4 S_{ab} \, ,
\end{equation}
where $S_{ab}$ is the surface stress tensor. In the case of pure tension branes, the surface stress tensor reads $S_{ab} = -\tau h_{ab}$.

In the C-metric, the surface $x = 0$ has the property that
\begin{equation}
    (K_{x})_{ab} = A (h_x)_{ab}
\end{equation}
while the surface $y = 0$ has
\begin{equation}
    (K_{y})_{ab} = A \sqrt{\lambda} (h_y)_{ab} \, .
\end{equation}
Since we are working throughout with two-sided branes, the extrinsic curvature on the other side of each brane is the negative of the extrinsic curvature computed above. Thus, for a pure tension brane with $K_{ab}=\alpha h_{ab}$, we have
\begin{equation}
[K_{ab}] = 2\alpha h_{ab} \, , \qquad [K]=6\alpha \, ,
\end{equation}
and therefore
\begin{equation}
\left[K_{ab}\right] - h_{ab} \left[K\right] = -4\alpha h_{ab} = -8\pi G_4 \tau\, h_{ab} \, .
\end{equation}
This gives
\begin{equation}
    \tau  = \frac{\alpha}{2 \pi  G_4} \, .
\end{equation}
Therefore, the tensions for the $x = 0$ and $y = 0$ branes are
\begin{equation}
    \tau_x = \frac{A}{2 \pi  G_4} \, ,\qquad \tau_y = \frac{A \sqrt{\lambda}}{2 \pi  G_4} \, .
\end{equation}

Using the formulas from above, we obtain
\begin{equation}
    \int_{x = 0} {\rm d}^3 x  \sqrt{h_x} \tau_x = \frac{\Delta \tau \Delta \phi}{4 \pi G_4 A^2} \left[\frac{1}{\epsilon^2}-\frac{1}{y_-^2} \right]
\end{equation}
and
\begin{equation}
    \int_{y = 0} {\rm d}^3 x  \sqrt{h_y} \tau_y = \frac{\lambda \Delta \tau \Delta \phi}{4 \pi G_4 A^2} \left[\frac{1}{\epsilon^2}-\frac{1}{x_+^2}\right] \, .
\end{equation}

\subsection*{Joint terms $J_{xz}$ and $J_{yz}$}

The region of Euclidean spacetime on which we are evaluating the action is piecewise smooth. It has corners where the $x = 0$ and $z = \epsilon$ surfaces meet, and also where the $y = 0$ and $z = \epsilon$ surfaces meet. For a well-posed variational problem, we must add Hayward joint terms~\cite{Hayward:1993my}:
\begin{equation}
I_{J_{xz}} = \int_{J_{xz}} {\rm d}^2 x \sqrt{\sigma_{xz}} \left(\eta_{xz} - \eta^{(0)}_{xz} \right) \, ,
\qquad
I_{J_{yz}} = \int_{J_{yz}} {\rm d}^2 x \sqrt{\sigma_{yz}} \left(\eta_{yz} - \eta^{(0)}_{yz} \right) \, .
\end{equation}
Here
\begin{equation}
\eta_{xz} = \arccos \left(n_{z} \cdot n_{x} \right) \, , \qquad
\eta_{yz} = \arccos \left(n_{z} \cdot n_{y} \right) \, ,
\end{equation}
and the integrals are performed over the two-dimensional spaces orthogonal to both surfaces.

Using the expressions from earlier, we obtain
\begin{equation}
    \eta_{xz} = \arctan \left(\sqrt{-H(-\epsilon)} \right) \, , \qquad
    \sqrt{\sigma_{xz}} = \frac{\sqrt{-H(-\epsilon)}}{A^2 \epsilon^2} \, ,
\end{equation}
and
\begin{equation}
    \eta_{yz} = \arctan \left(\sqrt{\frac{G(\epsilon)}{\lambda}} \right) \, , \qquad
    \sqrt{\sigma_{yz}} = \frac{\sqrt{\lambda G(\epsilon)}}{A^2 \epsilon^2} \, .
\end{equation}
Therefore,
\begin{equation}
    I_{J_{xz}} = \frac{\Delta \tau \Delta \phi\sqrt{-H(-\epsilon)}}{A^2 \epsilon^2} \left[\arctan \left(\sqrt{-H(-\epsilon)} \right) - \eta^{(0)}_{xz} \right]
\end{equation}
and
\begin{equation}
    I_{J_{yz}} = \frac{\Delta \tau \Delta \phi\sqrt{\lambda G(\epsilon)}}{A^2 \epsilon^2} \left[\arctan \left(\sqrt{\frac{G(\epsilon)}{\lambda}} \right) - \eta^{(0)}_{yz} \right] \, .
\end{equation}

A natural choice of counterterm is to subtract the limiting $\epsilon=0$ angle for the $\mu=0$ AdS$_4$ background:
\begin{equation}
   \eta^{(0)}_{xz} = \arctan \sqrt{\lambda} \, , \qquad  \eta^{(0)}_{yz} = \arctan \frac{1}{\sqrt{\lambda}} \, .
\end{equation}
Combining the two corner terms, expanding for small $\epsilon$, and simplifying gives
\begin{equation}
    I_{J_{xz}}  + I_{J_{yz}}  = \frac{\Delta \tau\Delta \phi}{A^{2}}
\left[k\frac{1-\lambda}{2 (\lambda+1)}+\mathcal{O}(\epsilon)\right]\, .
\end{equation}

\subsection*{Total action}

Collecting all terms, the renormalized on-shell action is
\begin{equation}
I_E^{\rm ren}
=
\frac{\Delta \tau \Delta \phi}{8\pi G_4 A^2}
\left[
\frac{\lambda+1}{(x_+-y_-)^2}
-\frac{1}{x_+^2}
-\frac{\lambda}{y_-^2}
\right] \, .
\end{equation}
This result agrees with the much earlier computation of the action by Kudoh and Kurita for a two-brane configuration~\cite{Kudoh:2004ub}; see also~\cite{Panella:2024sor, Cao:2026jls}.

\bibliographystyle{JHEP-2}
\bibliography{bib.bib}

\end{document}